\newcommand{\DTN} {NiCl$_2$$\cdot$4SC(NH$_2$)$_2$}
\newcommand{\DTNX} {Ni(Cl$_{1-x}$Br$_x$)$_2$$\cdot$4SC(NH$_2$)$_2$}

\newcommand{\aver}[1]{\left\langle #1 \right\rangle}

\documentclass[aps,prl,reprint,showpacs,longbibliography,superscriptaddress, citeonline]{revtex4-1}
\usepackage{amsmath,amssymb,bm}
\usepackage{graphicx}
\usepackage{xspace}
\usepackage{latexsym}
\usepackage{color}
\usepackage{hyperref}
\usepackage{placeins}

\begin{document}
\title{Quantum criticality in a three-dimensional spin system at zero field and pressure}

\author{K.~Yu.~Povarov}
    \email{povarovk@phys.ethz.ch}
    \affiliation{Neutron Scattering and Magnetism, Laboratory for Solid State Physics, ETH Z\"{u}rich, Switzerland}
    \homepage{http://www.neutron.ethz.ch/}

\author{A.~Mannig}
    \affiliation{Neutron Scattering and Magnetism, Laboratory for Solid State Physics, ETH Z\"{u}rich, Switzerland}

\author{G.~Perren}
    \affiliation{Neutron Scattering and Magnetism, Laboratory for Solid State Physics, ETH Z\"{u}rich, Switzerland}

\author{J.~S.~M\"{o}ller}
    \affiliation{Neutron Scattering and Magnetism, Laboratory for Solid State Physics, ETH Z\"{u}rich, Switzerland}

\author{E.~Wulf}
    \affiliation{Neutron Scattering and Magnetism, Laboratory for Solid State Physics, ETH Z\"{u}rich, Switzerland}

\author{J.~Ollivier}
    \affiliation{Institut Laue-Langevin, 6 rue Jules Horowitz, 38042 Grenoble, France}

\author{A.~Zheludev}
    \affiliation{Neutron Scattering and Magnetism, Laboratory for Solid State Physics, ETH Z\"{u}rich, Switzerland}

\begin{abstract}
We report on the spontaneous appearance of antiferromagnetic order
in a model gapped quantum paramagnet
Ni(Cl$_{1-x}$Br$_x$)$_2$$\cdot$4SC(NH$_2$)$_2$ induced by a change
in bromine concentration $x$. This transition is qualitatively
similar to a  $z = 1$ magnetic quantum critical point. However, the
observed critical scaling of thermodynamic and magnetic properties
has rather unusual  critical exponents.
\end{abstract}

\date{\today}
\maketitle

Magnetic insulators with their short-range and easily customizable
interactions are ideal models of various types of quantum critical
points (QCPs) and phase transitions
\cite{Sachdev_2011_QPTBook,Vojta_RevProgPhys_2003_QCPreview}. A
paradigmatic example is the transition between a gapped quantum
paramagnet and the semiclassical ordered N\'{e}el
phase~\cite{Sachdev_Science_2000_QCPs,*Sachdev_NPhys_2008_QCPs}. In
practice, such QCPs are typically induced by an external magnetic
field~\cite{Giamarchi_NatPhys_2008_BECreview,ColdeaTennant_Science_2010_IsingE8}
or by continuously tuning the Hamiltonian parameters. In real
quantum magnetic materials, the latter can sometimes be achieved by
applying hydrostatic  pressure
~\cite{Ruegg_PRL_2008_PindTlCuCl,Thede_PRL_2014_uSRPHCC,*PerrenMoeller_PRB_2015_PHCCpressurized}.
These zero field QCPs are rare but of a particular interest
\cite{Merchant_NatPhys_2014_PindTlCuCl,ScammellSushkov_PRB_2015_AsympFreedom}.
They break a continuous spin rotation symmetry and have a dynamical
critical exponent $z=1$. As a result, their properties are quite
distinct from those of the more familiar field induced
Bose--Einstein magnon
condensation~\cite{Giamarchi_NatPhys_2008_BECreview} or Ising-type
transitions
\cite{ColdeaTennant_Science_2010_IsingE8,HaelgHuvonen_PRB_2015_NTENPscaling}.
Unfortunately, the experimental necessity of using bulky pressure
cells for reaching the quantum critical points in model magnets
precludes many measurements needed to probe critical behavior,
universality, and scaling laws.

In the present Rapid Communication we report a three-dimensional
$XY$ spin system that is {\em quantum critical in zero applied field
and ambient pressure}. Our target material is the well known spin
gap compound \DTNX, which we tune to criticality by varying chemical
composition. We show that while the $x=0$ parent compound is gapped,
increasing Br concentration $x$ leads to a decrease and eventual
closure of the spin gap, followed by the appearance of magnetic long
range order with a gapless linear excitation spectrum. We then focus
on the material very close to the critical Br content $x_c$ and
study the critical properties and scaling at the apparent QCP.

%=======================================================================

\begin{figure}
  \includegraphics[width=0.5\textwidth]{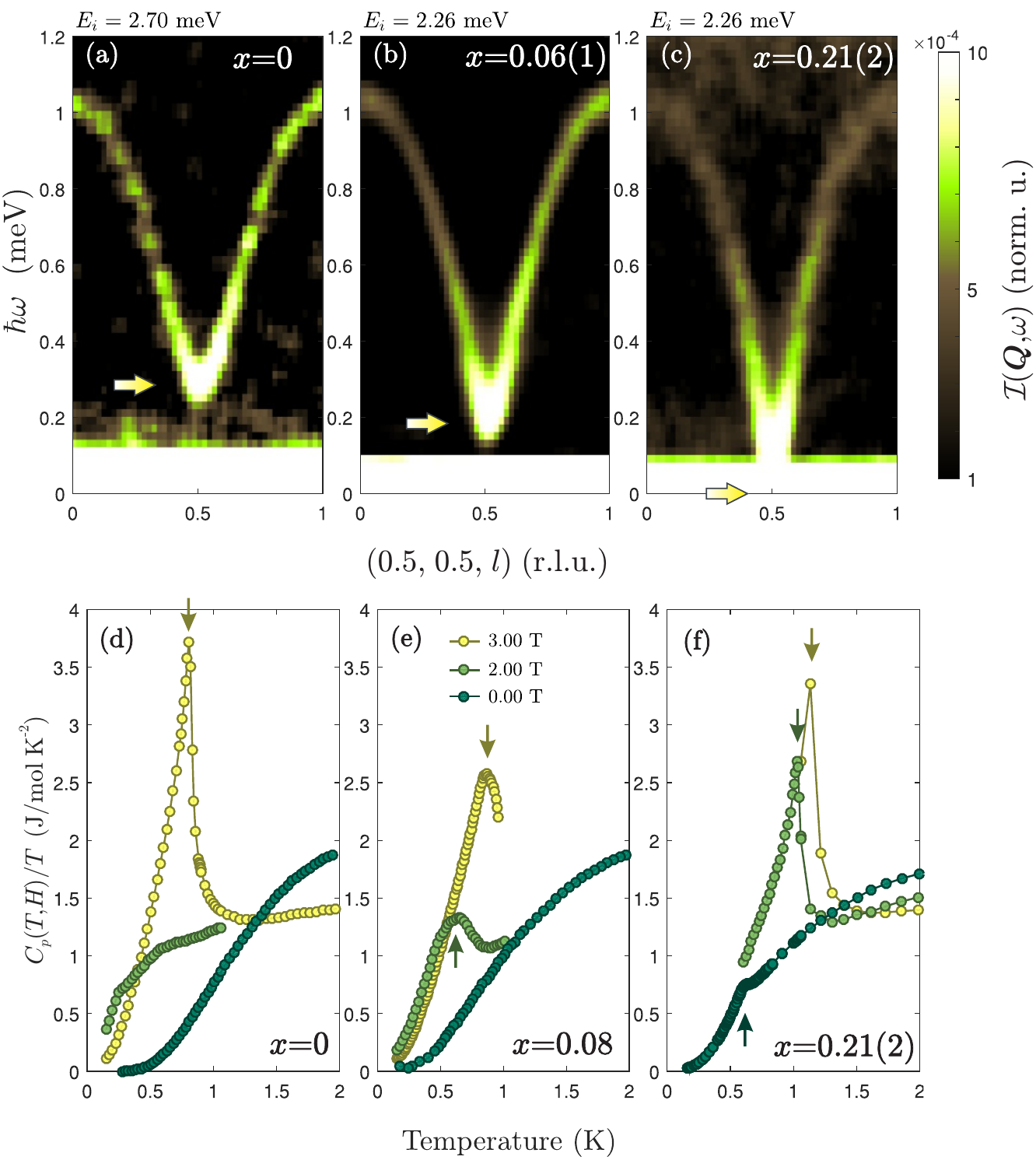}\\
  \caption{(a)--(c) Time of flight neutron scattering  spectra showing magnetic excitations in \DTNX\ with various Br concentration
$x$,
  traced along $\mathbf{Q}=(0.5,~0.5,~l)$ reciprocal space direction. Arrows indicate the corresponding gap values. The data were
  taken on the   IN5 instrument at temperatures of about $100$~mK. The incident
  neutron energy for each data set is indicated separately. The data in (b) are from
  Ref.~\cite{Povarov_PRB_2015_DTNXIN5}. (d)--(f) Specific heat of \DTNX\ samples with various $x$
  in magnetic field of $0$, $2$, and $3$~T. The arrows show the corresponding ordering temperatures, defined
  via the specific heat anomalies. The zero-field curve for $x=0.08$ is reproduced from Ref.~\cite{YuYin_Nat_2012_DTNboseglass}.}
  \label{FIG:overcrit}
\end{figure}

\begin{figure}
  \includegraphics[width=0.5\textwidth]{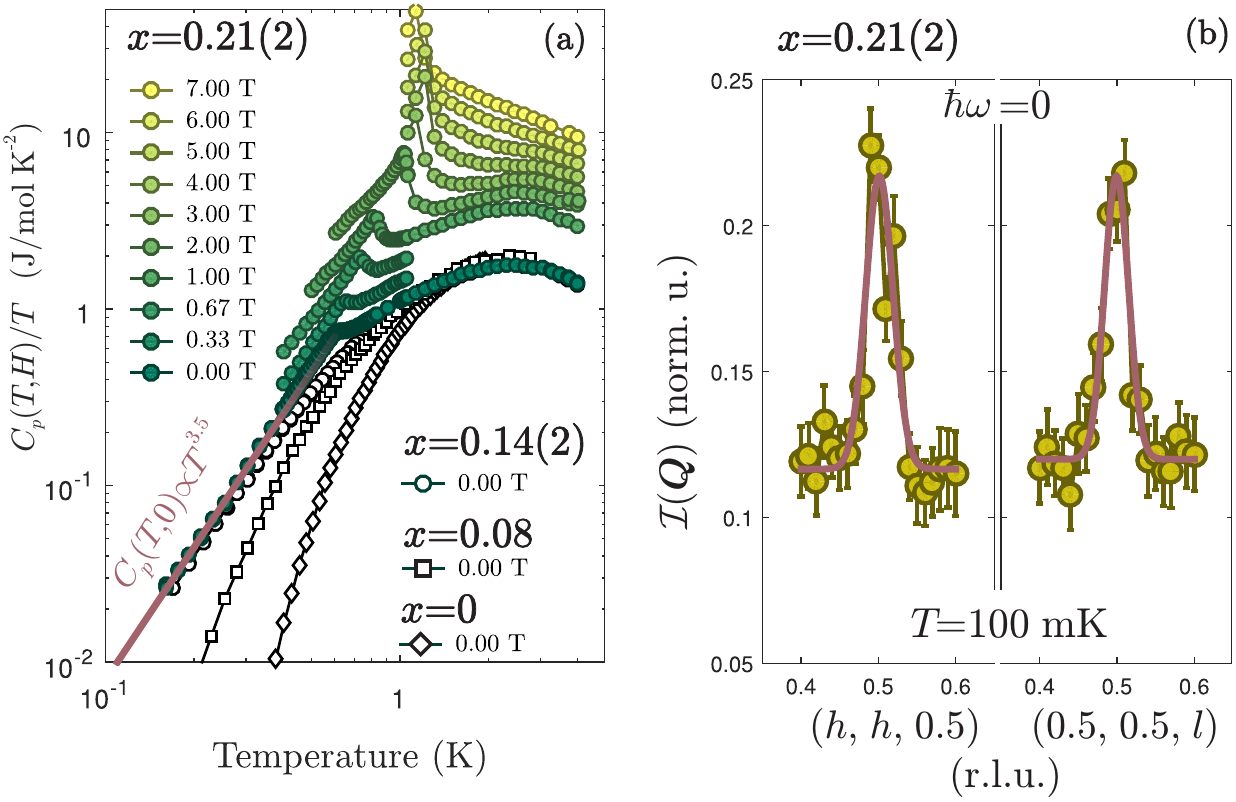}\\
  \caption{(a) Specific heat in ``overcritical'' $x=0.21(2)$ DTNX sample, given in comparison with zero-field data
  for ``critical'' $x=0.14(2)$, ``undercritical'' $x=0.08$, and clean $x=0$ samples. For nonzero $H$ a vertical
  offset is introduced for the data: $C_p$ of each curve is multiplied by a factor of $1.3^{N}$, where $N$ enumerates the curves
  with respect to ascending $H$. The field is
  applied along the high symmetry $c$
  direction. The zero-field curve for $x=0.08$ is reproduced from Ref.~\cite{YuYin_Nat_2012_DTNboseglass}.
  (b) The magnetic Bragg peak at $(0.5,~0.5,~0.5)$~r.l.u. in the
  $x=0.21(2)$ sample measured at low temperature. Neutron scattering intensity
  $\mathcal{I}(\mathbf{Q})$ for $\mathbf{Q}$ along and transverse to
  the $c$
  direction is shown in the left and right subpanels
  correspondingly. The points are the experimental data; solid lines
  are Gaussian fits.}
  \label{FIG:overcrit2}
\end{figure}

The parent compound \DTN\ has been extensively studied as a
prototypical spin gap material, particularly in the context of its
field-induced quantum phase
transitions~\cite{Zapf_PRL_2006_BECinDTN,*Zvyagin_PRL_2007_ESRinDTN,*YinXia_PRL_2008_DTNcritical,*BlinderDupont_PRB_2017_DTNNMR}.
A detailed description of the crystal structure, effective magnetic
Hamiltonian, and the role of Br substitution on the Cl site can be
found, for example, in our previous
work~\cite{Povarov_PRB_2015_DTNXIN5}. The magnetism is due to  $S=1$
Ni$^{2+}$ ions arranged on a tetragonal lattice, space group $I4$
and lattice parameters $a=9.56$ and $c=8.98$~\AA. The key energy
scales are the easy-plane single ion anisotropy $D=0.7$~meV, the
Heisenberg nearest neighbor exchange interactions along the unique
tetragonal axis $J_{c}=0.15$~meV, and weaker coupling
$J_{a}\simeq0.1J_{c}$ within each tetragonal plane. The planar
anisotropy term favors a nonmagnetic $S^{z}=0$ state for each spin,
while interactions favor N\'{e}el long range order. In the parent
compound the anisotropy term dominates, so that the ground state is
a nonmagnetic singlet. The lowest excitations are a highly
dispersive $S^{z}=\pm1$ doublet with an energy gap $\Delta=0.3$~meV.
These gapped excitations are readily seen by neutron spectroscopy
\cite{Zapf_PRL_2006_BECinDTN}. For reference, in
Fig.~\ref{FIG:overcrit}(a) we show our own data. These were taken
using two fully deuterated single crystal samples with total mass of
300~mg on the IN5 spectrometer at
ILL~\cite{OllivierMutka_JPSJ_2011_IN5}, using a $^3$He-$^4$He
dilution cryostat and neutrons with fixed incident energy of
2.7~meV. The chopper rotation speed for this measurement was set at
$12000$~rpm. The false color plot shows scattering intensity versus
energy transfer $\hbar\omega$ and momentum transfer along
$\mathbf{Q}=(0.5,~0.5,~l)$ reciprocal space rod, the
antiferromagnetic zone center $(0.5,~0.5,0.5)$ being the global
dispersion minimum where the gap is observed directly.

The Br-substituted material \DTNX\ has attracted a great deal of
interest in the context of effects of disorder on the field-induced
quantum phase transitions
\cite{YuYin_Nat_2012_DTNboseglass,ZheludevRoscilde_CRPhysique_2013_ReviewDirtyBosons,DupontCapponi_PRL_2017_DTNX2BEC}
and magnetic
excitations~\cite{Povarov_PRB_2015_DTNXIN5,OrlovaBlinder_PRL_2017_DTNXNMR}.
However, as was clearly shown by neutron spectroscopy studies, the
most obvious effect of Br substitution can be understood simply in
terms of its influence on {\em average} exchange and anisotropy
constants~\cite{Povarov_PRB_2015_DTNXIN5}. Specifically, increasing
$x$ decreases the $D/J_c$ ratio and thereby leads to a reduction of
the energy gap. For $x=0.06$ the measured spin excitation spectrum
is shown in Fig.~\ref{FIG:overcrit}(b) and corresponds to
$\Delta=0.2$~meV. Based on a simple linear extrapolation, our
previous analysis of the concentration dependence of the gap energy
predicted that it will be driven to zero somewhere around $x=0.2$
~\cite{Povarov_PRB_2015_DTNXIN5}.

The central result of the present study is that this indeed is the
case. Fully deuterated \DTNX\ single crystals with $x=0.21(2)$  were
grown  from solution following the procedure outlined in
Ref.~\cite{Wulf_2015_PhDthesis}. Single crystal x-ray diffraction
carried out on an Apex-II Bruker diffractometer confirmed the
structure to be almost identical to that of the parent material,
with lattice parameters $a=9.66$ and $c=9.08$~\AA\ and a homogeneous
Br distribution. Inelastic neutron scattering data for the sample
consisting of two coaligned $x=0.21(2)$ crystals with a total mass
of $200$~mg was collected on the IN5 spectrometer under the
conditions, identical to the experiment of
Ref.~\cite{Povarov_PRB_2015_DTNXIN5}: temperature of about $100$~mK
and neutron beam incident energy $E_{\text{i}}=2.26$~meV (with the
choppers rotating at $5000$~rpm). The resulting data are shown in
Fig.~\ref{FIG:overcrit}(c). In contrast to the other two spectra
measured for lower Br content, for $x=0.21(2)$ the spectrum is
gapless and approximately linear in the vicinity of the
antiferromagnetic zone center~\footnote{A detailed analysis of this
data will be given elsewhere (Mannig, Povarov \emph{et al.}, in
preparation).}.

Not only is the $x=0.21(2)$ system gapless, it is also magnetically
ordered at low temperatures. The magnetic phase transition was
detected by specific heat measurements performed on a Quantum Design
PPMS with a dilution cryostat insert. To illustrate the ordering
evolution with bromine content increase we compare the typical
specific heat curves from various samples [clean, $x=0.08$, and
$x=0.21(2)$] shown in Figs.~\ref{FIG:overcrit}(d)--(f). For an
applied magnetic field $H=3$~T, which exceeds the critical ordering
field $H_{c}^{x=0}=2.1$~T in the parent compound, all three samples
show clear lambda anomalies corresponding to the onset of long-range
order. The second field shown, $H=2$~T, is below $H_{c}^{x=0}$, but
above $H_{c}^{x=0.08}=1.1$~T for the $x=0.08$ sample. Under these
conditions, the lambda anomaly is present only in the two sample
with higher Br content. Finally, at zero applied field, only the
$x=0.21(2)$ sample still shows a tiny but distinct signature of a
phase transition at $T_{N}\simeq0.64$~K, while the two other samples
remain paramagnetic. The specific heat measurements also confirm the
gapless nature of the spectrum in the $x=0.21(2)$ sample. The
log-log specific heat curves shown in Fig.~\ref{FIG:overcrit2}(a)
contrast the activated (gapped) behavior for $x=0$ and $x=0.08$ with
a power law (gapless) trend for $x=0.21(2)$. For later reference,
take note of the power law exponent $\alpha=3.5\pm0.05$ fitted in
the temperature range 0.15--0.5~K.

Magnetic ordering in the $x=0.21(2)$ sample is also confirmed by
neutron diffraction. In the same IN5 data set as mentioned above, at
100~mK magnetic Bragg peaks are found at $(0.5,~0.5,~0.5)$
reciprocal space positions [symbols in Fig.~\ref{FIG:overcrit2}(b)].
These peaks are resolution-limited, as deduced from Gaussian fits to
intensity profiles measured along different directions (solid
lines). An analysis of their
intensities~\cite{Squires_2012_Neutronbook} (see Supplemental
Maaterial for details) allows us to make a crude estimate of the
ordered moment $m\simeq0.3\mu_{B}$, which is very small compared to
the classical saturation value $2\mu_B$ for $S=1$. Note that in zero
applied field the ordering vector in the $x=0.21(2)$ sample is the
same as in the field-induced ordered phase of the $x=0$ parent
compound~\cite{Tsyrulin_JPCM_2013_DTNneutrons,WulfHuvonen_PRB_2013_DTNXdiffraction}.

\begin{figure}
  \includegraphics[width=0.4\textwidth]{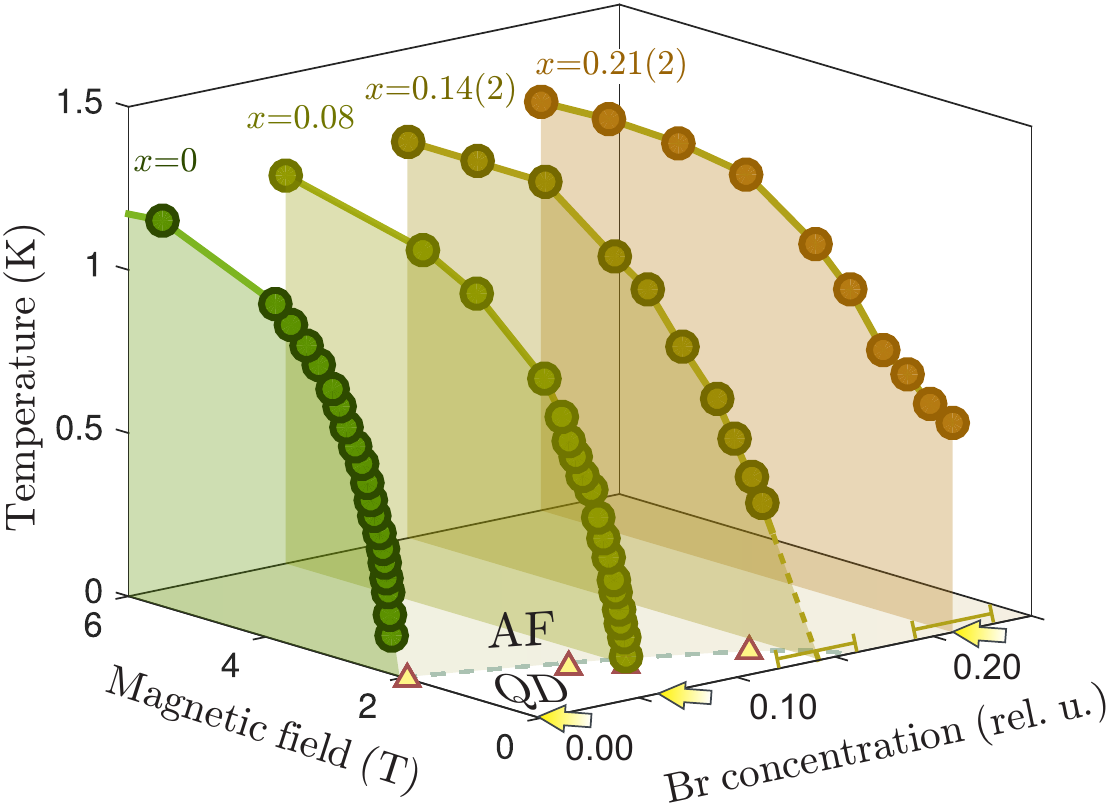}\\
  \caption{Phase diagrams of DTNX with various concentration of bromine $x$ in the magnetic field along the $c$ direction. Solid circles are the specific heat data,
  solid lines are guides to the eye (for $x=0$ and $0.08$ the data is from Refs.~\cite{Zapf_PRL_2006_BECinDTN,YuYin_Nat_2012_DTNboseglass}).
  Solid triangles are the $H_{c}(x)$ values, known from the literature~\cite{Zapf_PRL_2006_BECinDTN,YuYin_Nat_2012_DTNboseglass,WulfHuvonen_PRB_2013_DTNXdiffraction,Povarov_PRB_2015_DTNXIN5}.
  The dashed line marks the \textit{tentative} boundary between the antiferromagnetically ordered (AF) and quantum
  disordered (QD)
  states. Concentration errorbars for the present study are also
  shown. Arrows correspond to the ``points'' on the phase diagram, at which the neutron data
  in Fig.~\ref{FIG:overcrit} was measured.
  }
  \label{FIG:Diagrams}
\end{figure}

\begin{figure}
  \includegraphics[width=0.5\textwidth]{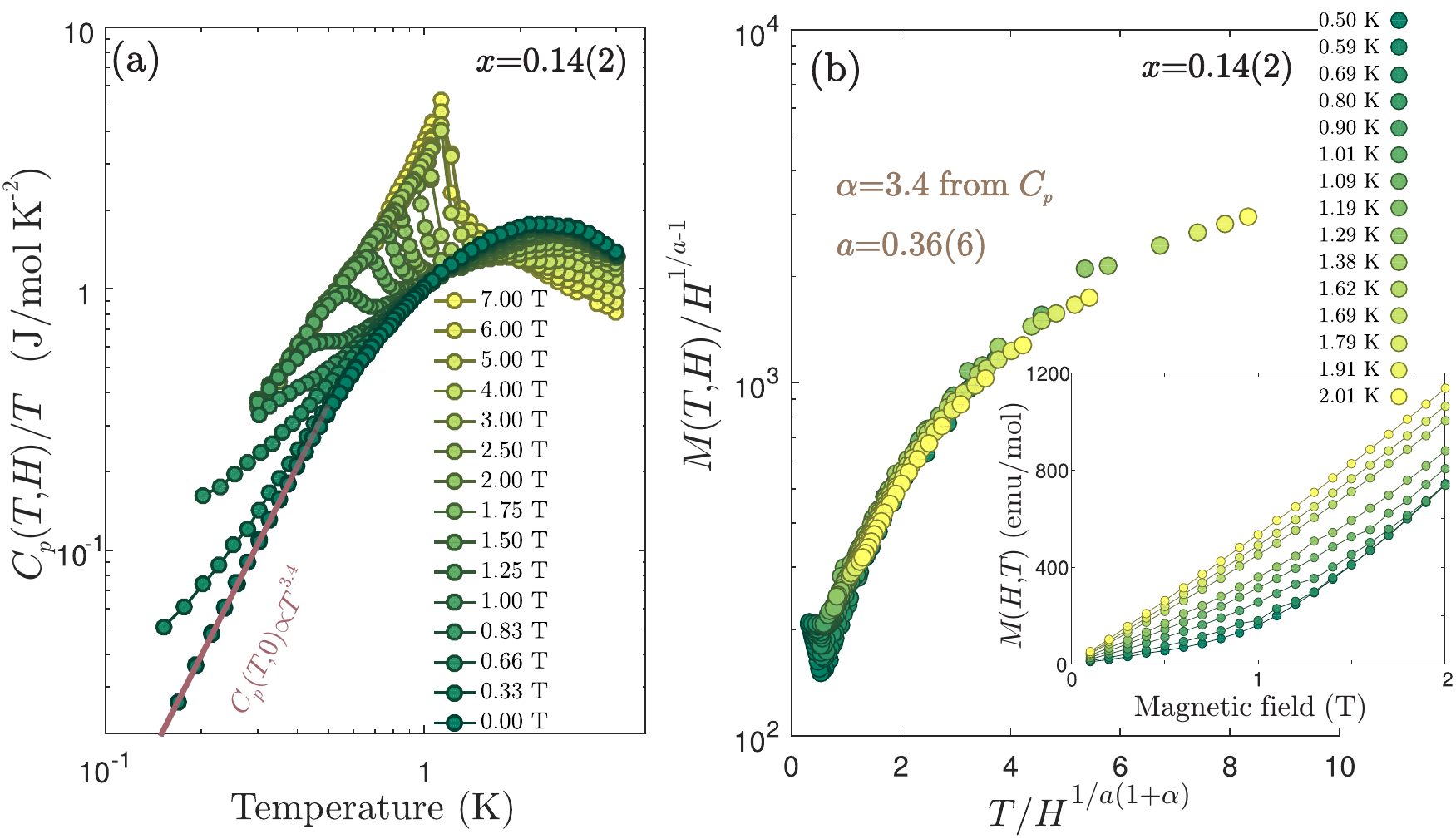}\\
  \caption{
  Thermodynamic properties and scaling in the ``critical''
  $x\simeq14$\% DTNX sample.
  (a) Specific heat in various magnetic fields. At $H=0$ a
  low-temperature limit with the specific heat proportional to
  $T^{3.4}$ is found in the absence of ordering.
  (b) Main
  panel: magnetization data plotted in scaled variables. The
  parameters optimizing the scaling are given in the plot.
  Inset: original magnetization curves at various temperatures.
In all measurements the magnetic field is applied along the high
symmetry $c$
  direction.
  }
  \label{FIG:Mscaling}
\end{figure}

The combined phase diagram for \DTNX\ based on our data and those
found in literature, is shown in Fig.~\ref{FIG:Diagrams}. The
transition temperatures in this graph were determined from the
specific heat anomalies, such as those shown in
Figs.~\ref{FIG:overcrit2} and  \ref{FIG:Mscaling}. Based on the
available data it is not easy to accurately pinpoint the critical
concentration $x_c$ at which long range order appears in zero
applied field. However, from the known $x$ dependence of the
critical field at $T\rightarrow 0$ (triangles in
Fig.~\ref{FIG:Diagrams}) we can crudely estimate
$x_{c}\simeq0.16$~\footnote{The value of $x_c\simeq0.16$ is also
consistent with available microscopic models for
DTNX~\cite{DupontCapponi_PRL_2017_DTNX2BEC}: It corresponds to the
percolation threshold of nickel sites, affected by neighboring
bromine.}. Of all samples synthesized for the present study, the one
with the closest Br content has $x=0.14(2)$. Indeed, this material
appears to be on the verge of spontaneous ordering. As shown in
Fig.~\ref{FIG:Mscaling}, the lambda anomalies in specific heat  are
unresolved in fields below $0.8$~T (or equivalently, temperatures
below $300$~mK). Nevertheless, down to at least $0.3$~T, the heat
capacity curves contain precursors of long-range ordering, namely
upturns in $C_{p}(T)/T$ at $T\rightarrow 0$. No such precursors are
present in zero applied field. Instead, the specific heat at $H=0$
seemingly follows the same power law
\begin{equation}
C_{p}(T)\propto T^{\alpha}\label{EQ:Cv}
\end{equation}
with $\alpha=3.4\pm0.15$, within the error coinciding with the value
of $3.5$ found in a well-ordered $x=0.21(2)$ sample. In fact, as
Fig.~\ref{FIG:overcrit2}(a) directly shows, in the low-temperature
limit $C_{p}(T)$ for $x=0.21(2)$ and $x=0.14(2)$ samples converge to
the same trend. This strongly suggests that for $x=0.14(2)$ the
low-energy excitations are also gapless.

From the experimental point of view, the $x=0.14(2)$ sample thus
appears to be at or very close to a quantum critical point. With
this assumption, we can learn more about the underlying physics by
checking the scaling of another readily accessible quantity, namely
magnetization in the magnetic field, applied along the anisotropy
axis ($c$ direction). Usage of the iQuantum $^3$He insert for the
Quantum Design MPMS SQUID magnetometer allows us to perform the
magnetometry measurements in the sub-Kelvin regime. Typical
magnetization curves in the $x=0.14(2)$ sample at different
temperatures are shown in the inset of Fig.~\ref{FIG:Mscaling}(b).
Criticality implies that the free energy of the system obeys the
scaling
relation~\cite{Sachdev_2011_QPTBook,Vojta_RevProgPhys_2003_QCPreview}:
\begin{equation}\label{EQ:scalingmain}
    F(T,H)=\lambda^{-1}\mathcal{F}(\lambda^{\frac{1}{1+\alpha}}T,\lambda^{a}H),
\end{equation}

for any positive $\lambda$, with scaling exponents $a$ and $\alpha$,
the latter being the same as in Eq.~(\ref{EQ:Cv}). For magnetization
this gives the following scaling form:

\begin{equation}
    M(T,H)/H^{\frac{1-a}{a}}=\mathcal{M}(T/H^{\frac{1}{a(1+\alpha)}}),
\end{equation}

where $\mathcal{M}(x)$ is an \textit{a priori} unknown scaling
function of a single variable. With a proper choice of the exponent
$a$, all measured  $M(T,H)/H^{\frac{1-a}{a}}$ data should collapse
onto a single curve when plotted vs $T/H^{\frac{1}{a(1+\alpha)}}$.
Fixing $\alpha\simeq3.4$ based on the calorimetric measurements, to
find the optimal value of $a$ we minimize the empirical ``data
overlap'' function, as it was done in a number of scaling studies
before~\cite{PovarovSchmidiger_PRB_2015_DimpyScaling,HaelgHuvonen_PRB_2015_NTENPscaling,HaelgHuvonen_PRB_2015_Dioxanescaling}.
With $a$ as the only adjustable parameter, an excellent data
collapse [main panel of Fig.~\ref{FIG:Mscaling}(b)]is obtained at
$a=0.36(6)$.

If we assume that the transition in \DTNX\ can be fully understood
in terms of the variation of  average Hamiltonian parameters with Br
concentration, we are dealing with a $z=1$ QCP in $d=3$ dimensions
with a spontaneous breaking of $O(2)$ symmetry
~\cite{ZhangWierschem_PRB_2013_DTNdispersion}. Interestingly, the
observed critical exponents do {\em not} agree with this model.
There the dynamical exponent $z=1$, and hyperscaling immediately
gives $\alpha=d/z=3$.  There could be a number of trivial
explanations for such a discrepancy, such as possibly insufficient
dynamic range in our experiments, or an inhomogeneous stress
distribution in the samples that is known to have a rather drastic
effect on the field induced transition in the parent compound
\cite{WulfHuvonen_PRB_2015_DTNIntrinsicBroadening}. Another factor
may be a violation of hyperscaling due to dangerously irrelevant
magnon-magnon
interaction~\cite{Fisher_PRB_1989_Boseglass,YuMiclea_PRB_2012_DTNXexponents}.
The latter may be of special importance as $z=1$ $d=3$ QCP lies
exactly at the upper critical dimension of a phase transition that,
as in our case, breaks $O(2)$ symmetry. The most interesting
interpretation though has to do with disorder, which to this point
we deliberately left out of the discussion. The ``clean'' $O(2)$
transition has the mean-field correlation length critical exponent
$\nu=1/2$~\cite{ZhangWierschem_PRB_2013_DTNdispersion} not
satisfying the necessary condition $d\nu>2$ under which the disorder
vanishes on large scales. This so-called Harris
criterion~\cite{Harris_JPhysC_1974_HarrisCriter} implies that
disorder in \DTNX\ {\em is} relevant, although whether it has any
measurable effect on the criticality of the $z=2$ field-induced
transitions is controversial
\cite{ZhiyuanProkofev_PRL_2014_BGexponent,WulfHuvonen_PRB_2015_DTNIntrinsicBroadening}.
In our case, the $z=1$ transition is expected to be more
susceptible. In fact, the QCP is not expected to survive in the
presence of disorder, which gives rise to a strongly inhomogeneous
weakly ordered Griffiths phase~\cite{Vojta_PRL_2013_InGap}. That
disorder must play a role in the concentration-induced transition in
\DTNX\ is also supported by the unusual specific heat power law in
the well ordered $x=0.21(2)$ sample. Indeed, the contribution of
linearly dispersive antiferromagnetic spin waves would simply
correspond to $\alpha=3$.

In any case, further experimental work and theoretical guidance will
be required to fully understand the new concentration-induced
transition and criticality in \DTNX. Fortunately, the transition
occurs in zero applied field and ambient pressure, which will enable
a host of future experiments.

%====================================================

\acknowledgments This work was supported by Swiss National Science
Foundation, Division II. We would like to thank Dr. S.~Gvasaliya
(ETH Z\"{u}rich) for assistance with the sample alignment for the
neutron experiment.

\bibliography{d:/The_Library}

\begin{thebibliography}{37}%
\makeatletter
\providecommand \@ifxundefined [1]{%
 \@ifx{#1\undefined}
}%
\providecommand \@ifnum [1]{%
 \ifnum #1\expandafter \@firstoftwo
 \else \expandafter \@secondoftwo
 \fi
}%
\providecommand \@ifx [1]{%
 \ifx #1\expandafter \@firstoftwo
 \else \expandafter \@secondoftwo
 \fi
}%
\providecommand \natexlab [1]{#1}%
\providecommand \enquote  [1]{``#1''}%
\providecommand \bibnamefont  [1]{#1}%
\providecommand \bibfnamefont [1]{#1}%
\providecommand \citenamefont [1]{#1}%
\providecommand \href@noop [0]{\@secondoftwo}%
\providecommand \href [0]{\begingroup \@sanitize@url \@href}%
\providecommand \@href[1]{\@@startlink{#1}\@@href}%
\providecommand \@@href[1]{\endgroup#1\@@endlink}%
\providecommand \@sanitize@url [0]{\catcode `\\12\catcode `\$12\catcode
  `\&12\catcode `\#12\catcode `\^12\catcode `\_12\catcode `\%12\relax}%
\providecommand \@@startlink[1]{}%
\providecommand \@@endlink[0]{}%
\providecommand \url  [0]{\begingroup\@sanitize@url \@url }%
\providecommand \@url [1]{\endgroup\@href {#1}{\urlprefix }}%
\providecommand \urlprefix  [0]{URL }%
\providecommand \Eprint [0]{\href }%
\providecommand \doibase [0]{http://dx.doi.org/}%
\providecommand \selectlanguage [0]{\@gobble}%
\providecommand \bibinfo  [0]{\@secondoftwo}%
\providecommand \bibfield  [0]{\@secondoftwo}%
\providecommand \translation [1]{[#1]}%
\providecommand \BibitemOpen [0]{}%
\providecommand \bibitemStop [0]{}%
\providecommand \bibitemNoStop [0]{.\EOS\space}%
\providecommand \EOS [0]{\spacefactor3000\relax}%
\providecommand \BibitemShut  [1]{\csname bibitem#1\endcsname}%
\let\auto@bib@innerbib\@empty
%</preamble>
\bibitem [{\citenamefont {Sachdev}(2011)}]{Sachdev_2011_QPTBook}%
  \BibitemOpen
  \bibfield  {author} {\bibinfo {author} {\bibfnamefont {S.}~\bibnamefont
  {Sachdev}},\ }\href
  {http://www.cambridge.org/ch/academic/subjects/physics/condensed-matter-physics-nanoscience-and-mesoscopic-physics/quantum-phase-transitions-2nd-edition?format=HB}
  {\emph {\bibinfo {title} {Quantum Phase Transitions}}}\ (\bibinfo
  {publisher} {Cambridge University Press, U.K.},\ \bibinfo {year}
  {2011})\BibitemShut {NoStop}%
\bibitem [{\citenamefont {Vojta}(2003)}]{Vojta_RevProgPhys_2003_QCPreview}%
  \BibitemOpen
  \bibfield  {author} {\bibinfo {author} {\bibfnamefont {M.}~\bibnamefont
  {Vojta}},\ }\bibfield  {title} {\enquote {\bibinfo {title} {Quantum phase
  transitions},}\ }\href {\doibase 10.1088/0034-4885/66/12/R01} {\bibfield
  {journal} {\bibinfo  {journal} {Rep. Prog. Phys.}\ }\textbf {\bibinfo
  {volume} {66}},\ \bibinfo {pages} {2069} (\bibinfo {year}
  {2003})}\BibitemShut {NoStop}%
\bibitem [{\citenamefont {Sachdev}(2000)}]{Sachdev_Science_2000_QCPs}%
  \BibitemOpen
  \bibfield  {author} {\bibinfo {author} {\bibfnamefont {S.}~\bibnamefont
  {Sachdev}},\ }\bibfield  {title} {\enquote {\bibinfo {title} {Quantum
  criticality: Competing ground states in low dimensions},}\ }\href {\doibase
  10.1126/science.288.5465.475} {\bibfield  {journal} {\bibinfo  {journal}
  {Science}\ }\textbf {\bibinfo {volume} {288}},\ \bibinfo {pages} {475}
  (\bibinfo {year} {2000})}\BibitemShut {NoStop}%
\bibitem [{\citenamefont {Sachdev}(2008)}]{Sachdev_NPhys_2008_QCPs}%
  \BibitemOpen
  \bibfield  {author} {\bibinfo {author} {\bibfnamefont {S.}~\bibnamefont
  {Sachdev}},\ }\bibfield  {title} {\enquote {\bibinfo {title} {Quantum
  magnetism and criticality},}\ }\href {\doibase 10.1038/nphys894} {\bibfield
  {journal} {\bibinfo  {journal} {Nat. Physics}\ }\textbf {\bibinfo {volume}
  {4}},\ \bibinfo {pages} {173} (\bibinfo {year} {2008})}\BibitemShut {NoStop}%
\bibitem [{\citenamefont {Giamarchi}\ \emph {et~al.}(2008)\citenamefont
  {Giamarchi}, \citenamefont {R{\"u}egg},\ and\ \citenamefont
  {Tchernyshyov}}]{Giamarchi_NatPhys_2008_BECreview}%
  \BibitemOpen
  \bibfield  {author} {\bibinfo {author} {\bibfnamefont {T.}~\bibnamefont
  {Giamarchi}}, \bibinfo {author} {\bibfnamefont {C.}~\bibnamefont
  {R{\"u}egg}}, \ and\ \bibinfo {author} {\bibfnamefont {O.}~\bibnamefont
  {Tchernyshyov}},\ }\bibfield  {title} {\enquote {\bibinfo {title}
  {{Bose-Einstein condensation in magnetic insulators}},}\ }\href {\doibase
  10.1038/nphys893} {\bibfield  {journal} {\bibinfo  {journal} {Nat. Physics}\
  }\textbf {\bibinfo {volume} {4}},\ \bibinfo {pages} {198} (\bibinfo {year}
  {2008})}\BibitemShut {NoStop}%
\bibitem [{\citenamefont {Coldea}\ \emph {et~al.}(2010)\citenamefont {Coldea},
  \citenamefont {Tennant}, \citenamefont {Wheeler}, \citenamefont {Wawrzynska},
  \citenamefont {Prabhakaran}, \citenamefont {Telling}, \citenamefont
  {Habicht}, \citenamefont {Smeibidl},\ and\ \citenamefont
  {Kiefer}}]{ColdeaTennant_Science_2010_IsingE8}%
  \BibitemOpen
  \bibfield  {author} {\bibinfo {author} {\bibfnamefont {R.}~\bibnamefont
  {Coldea}}, \bibinfo {author} {\bibfnamefont {D.~A.}\ \bibnamefont {Tennant}},
  \bibinfo {author} {\bibfnamefont {E.~M.}\ \bibnamefont {Wheeler}}, \bibinfo
  {author} {\bibfnamefont {E.}~\bibnamefont {Wawrzynska}}, \bibinfo {author}
  {\bibfnamefont {D.}~\bibnamefont {Prabhakaran}}, \bibinfo {author}
  {\bibfnamefont {M.}~\bibnamefont {Telling}}, \bibinfo {author} {\bibfnamefont
  {K.}~\bibnamefont {Habicht}}, \bibinfo {author} {\bibfnamefont
  {P.}~\bibnamefont {Smeibidl}}, \ and\ \bibinfo {author} {\bibfnamefont
  {K.}~\bibnamefont {Kiefer}},\ }\bibfield  {title} {\enquote {\bibinfo {title}
  {{Quantum Criticality in an Ising Chain: Experimental Evidence for Emergent
  E8 Symmetry}},}\ }\href {\doibase 10.1126/science.1180085} {\bibfield
  {journal} {\bibinfo  {journal} {Science}\ }\textbf {\bibinfo {volume}
  {327}},\ \bibinfo {pages} {177} (\bibinfo {year} {2010})}\BibitemShut
  {NoStop}%
\bibitem [{\citenamefont {R\"uegg}\ \emph {et~al.}(2008)\citenamefont
  {R\"uegg}, \citenamefont {Normand}, \citenamefont {Matsumoto}, \citenamefont
  {Furrer}, \citenamefont {McMorrow}, \citenamefont {Kr\"amer}, \citenamefont
  {G\"udel}, \citenamefont {Gvasaliya}, \citenamefont {Mutka},\ and\
  \citenamefont {Boehm}}]{Ruegg_PRL_2008_PindTlCuCl}%
  \BibitemOpen
  \bibfield  {author} {\bibinfo {author} {\bibfnamefont {Ch.}\ \bibnamefont
  {R\"uegg}}, \bibinfo {author} {\bibfnamefont {B.}~\bibnamefont {Normand}},
  \bibinfo {author} {\bibfnamefont {M.}~\bibnamefont {Matsumoto}}, \bibinfo
  {author} {\bibfnamefont {A.}~\bibnamefont {Furrer}}, \bibinfo {author}
  {\bibfnamefont {D.~F.}\ \bibnamefont {McMorrow}}, \bibinfo {author}
  {\bibfnamefont {K.~W.}\ \bibnamefont {Kr\"amer}}, \bibinfo {author}
  {\bibfnamefont {H.~U.}\ \bibnamefont {G\"udel}}, \bibinfo {author}
  {\bibfnamefont {S.~N.}\ \bibnamefont {Gvasaliya}}, \bibinfo {author}
  {\bibfnamefont {H.}~\bibnamefont {Mutka}}, \ and\ \bibinfo {author}
  {\bibfnamefont {M.}~\bibnamefont {Boehm}},\ }\bibfield  {title} {\enquote
  {\bibinfo {title} {{Quantum Magnets under Pressure: Controlling Elementary
  Excitations in ${\mathrm{TlCuCl}}_{3}$}},}\ }\href {\doibase
  10.1103/PhysRevLett.100.205701} {\bibfield  {journal} {\bibinfo  {journal}
  {Phys. Rev. Lett.}\ }\textbf {\bibinfo {volume} {100}},\ \bibinfo {pages}
  {205701} (\bibinfo {year} {2008})}\BibitemShut {NoStop}%
\bibitem [{\citenamefont {Thede}\ \emph {et~al.}(2014)\citenamefont {Thede},
  \citenamefont {Mannig}, \citenamefont {M\aa{}nsson}, \citenamefont
  {H\"uvonen}, \citenamefont {Khasanov}, \citenamefont {Morenzoni},\ and\
  \citenamefont {Zheludev}}]{Thede_PRL_2014_uSRPHCC}%
  \BibitemOpen
  \bibfield  {author} {\bibinfo {author} {\bibfnamefont {M.}~\bibnamefont
  {Thede}}, \bibinfo {author} {\bibfnamefont {A.}~\bibnamefont {Mannig}},
  \bibinfo {author} {\bibfnamefont {M.}~\bibnamefont {M\aa{}nsson}}, \bibinfo
  {author} {\bibfnamefont {D.}~\bibnamefont {H\"uvonen}}, \bibinfo {author}
  {\bibfnamefont {R.}~\bibnamefont {Khasanov}}, \bibinfo {author}
  {\bibfnamefont {E.}~\bibnamefont {Morenzoni}}, \ and\ \bibinfo {author}
  {\bibfnamefont {A.}~\bibnamefont {Zheludev}},\ }\bibfield  {title} {\enquote
  {\bibinfo {title} {Pressure-induced quantum critical and multicritical points
  in a frustrated spin liquid},}\ }\href {\doibase
  10.1103/PhysRevLett.112.087204} {\bibfield  {journal} {\bibinfo  {journal}
  {Phys. Rev. Lett.}\ }\textbf {\bibinfo {volume} {112}},\ \bibinfo {pages}
  {087204} (\bibinfo {year} {2014})}\BibitemShut {NoStop}%
\bibitem [{\citenamefont {Perren}\ \emph {et~al.}(2015)\citenamefont {Perren},
  \citenamefont {M\"oller}, \citenamefont {H\"uvonen}, \citenamefont
  {Podlesnyak},\ and\ \citenamefont
  {Zheludev}}]{PerrenMoeller_PRB_2015_PHCCpressurized}%
  \BibitemOpen
  \bibfield  {author} {\bibinfo {author} {\bibfnamefont {G.}~\bibnamefont
  {Perren}}, \bibinfo {author} {\bibfnamefont {J.~S.}\ \bibnamefont
  {M\"oller}}, \bibinfo {author} {\bibfnamefont {D.}~\bibnamefont {H\"uvonen}},
  \bibinfo {author} {\bibfnamefont {A.~A.}\ \bibnamefont {Podlesnyak}}, \ and\
  \bibinfo {author} {\bibfnamefont {A.}~\bibnamefont {Zheludev}},\ }\bibfield
  {title} {\enquote {\bibinfo {title} {{Spin dynamics in pressure-induced
  magnetically ordered phases in
  $({\mathrm{C}}_{4}{\mathrm{H}}_{12}{\mathrm{N}}_{2}){\mathrm{Cu}}_{2}{\mathrm{Cl}}_{6}$}},}\
  }\href {\doibase 10.1103/PhysRevB.92.054413} {\bibfield  {journal} {\bibinfo
  {journal} {Phys. Rev. B}\ }\textbf {\bibinfo {volume} {92}},\ \bibinfo
  {pages} {054413} (\bibinfo {year} {2015})}\BibitemShut {NoStop}%
\bibitem [{\citenamefont {Merchant}\ \emph {et~al.}(2008)\citenamefont
  {Merchant}, \citenamefont {Normand}, \citenamefont {Kr\"amer}, \citenamefont
  {Boehm}, \citenamefont {McMorrow},\ and\ \citenamefont
  {R\"uegg}}]{Merchant_NatPhys_2014_PindTlCuCl}%
  \BibitemOpen
  \bibfield  {author} {\bibinfo {author} {\bibfnamefont {P.}~\bibnamefont
  {Merchant}}, \bibinfo {author} {\bibfnamefont {B.}~\bibnamefont {Normand}},
  \bibinfo {author} {\bibfnamefont {K.~W.}\ \bibnamefont {Kr\"amer}}, \bibinfo
  {author} {\bibfnamefont {M.}~\bibnamefont {Boehm}}, \bibinfo {author}
  {\bibfnamefont {D.~F.}\ \bibnamefont {McMorrow}}, \ and\ \bibinfo {author}
  {\bibfnamefont {Ch.}\ \bibnamefont {R\"uegg}},\ }\bibfield  {title} {\enquote
  {\bibinfo {title} {Quantum and classical criticality in a dimerized quantum
  antiferromagnet},}\ }\href {\doibase 10.1038/nphys2902} {\bibfield  {journal}
  {\bibinfo  {journal} {Nat. Physics}\ }\textbf {\bibinfo {volume} {10}},\
  \bibinfo {pages} {373} (\bibinfo {year} {2008})}\BibitemShut {NoStop}%
\bibitem [{\citenamefont {Scammell}\ and\ \citenamefont
  {Sushkov}(2015)}]{ScammellSushkov_PRB_2015_AsympFreedom}%
  \BibitemOpen
  \bibfield  {author} {\bibinfo {author} {\bibfnamefont {H.~D.}\ \bibnamefont
  {Scammell}}\ and\ \bibinfo {author} {\bibfnamefont {O.~P.}\ \bibnamefont
  {Sushkov}},\ }\bibfield  {title} {\enquote {\bibinfo {title} {Asymptotic
  freedom in quantum magnets},}\ }\href {\doibase 10.1103/PhysRevB.92.220401}
  {\bibfield  {journal} {\bibinfo  {journal} {Phys. Rev. B}\ }\textbf {\bibinfo
  {volume} {92}},\ \bibinfo {pages} {220401} (\bibinfo {year}
  {2015})}\BibitemShut {NoStop}%
\bibitem [{\citenamefont {H\"alg}\ \emph
  {et~al.}(2015{\natexlab{a}})\citenamefont {H\"alg}, \citenamefont
  {H\"uvonen}, \citenamefont {Guidi}, \citenamefont {Quintero-Castro},
  \citenamefont {Boehm}, \citenamefont {Regnault}, \citenamefont {Hagiwara},\
  and\ \citenamefont {Zheludev}}]{HaelgHuvonen_PRB_2015_NTENPscaling}%
  \BibitemOpen
  \bibfield  {author} {\bibinfo {author} {\bibfnamefont {M.}~\bibnamefont
  {H\"alg}}, \bibinfo {author} {\bibfnamefont {D.}~\bibnamefont {H\"uvonen}},
  \bibinfo {author} {\bibfnamefont {T.}~\bibnamefont {Guidi}}, \bibinfo
  {author} {\bibfnamefont {D.~L.}\ \bibnamefont {Quintero-Castro}}, \bibinfo
  {author} {\bibfnamefont {M.}~\bibnamefont {Boehm}}, \bibinfo {author}
  {\bibfnamefont {L.~P.}\ \bibnamefont {Regnault}}, \bibinfo {author}
  {\bibfnamefont {M.}~\bibnamefont {Hagiwara}}, \ and\ \bibinfo {author}
  {\bibfnamefont {A.}~\bibnamefont {Zheludev}},\ }\bibfield  {title} {\enquote
  {\bibinfo {title} {{Finite-temperature scaling of spin correlations in an
  experimental realization of the one-dimensional Ising quantum critical
  point}},}\ }\href {\doibase 10.1103/PhysRevB.92.014412} {\bibfield  {journal}
  {\bibinfo  {journal} {Phys. Rev. B}\ }\textbf {\bibinfo {volume} {92}},\
  \bibinfo {pages} {014412} (\bibinfo {year} {2015}{\natexlab{a}})}\BibitemShut
  {NoStop}%
\bibitem [{\citenamefont {Povarov}\ \emph
  {et~al.}(2015{\natexlab{a}})\citenamefont {Povarov}, \citenamefont {Wulf},
  \citenamefont {H\"uvonen}, \citenamefont {Ollivier}, \citenamefont
  {Paduan-Filho},\ and\ \citenamefont {Zheludev}}]{Povarov_PRB_2015_DTNXIN5}%
  \BibitemOpen
  \bibfield  {author} {\bibinfo {author} {\bibfnamefont {{\relax K.
  Yu.}}~\bibnamefont {Povarov}}, \bibinfo {author} {\bibfnamefont
  {E.}~\bibnamefont {Wulf}}, \bibinfo {author} {\bibfnamefont {D.}~\bibnamefont
  {H\"uvonen}}, \bibinfo {author} {\bibfnamefont {J.}~\bibnamefont {Ollivier}},
  \bibinfo {author} {\bibfnamefont {A.}~\bibnamefont {Paduan-Filho}}, \ and\
  \bibinfo {author} {\bibfnamefont {A.}~\bibnamefont {Zheludev}},\ }\bibfield
  {title} {\enquote {\bibinfo {title} {{Dynamics of a bond-disordered $S=1$
  quantum magnet near $z=1$ criticality}},}\ }\href {\doibase
  10.1103/PhysRevB.92.024429} {\bibfield  {journal} {\bibinfo  {journal} {Phys.
  Rev. B}\ }\textbf {\bibinfo {volume} {92}},\ \bibinfo {pages} {024429}
  (\bibinfo {year} {2015}{\natexlab{a}})}\BibitemShut {NoStop}%
\bibitem [{\citenamefont {Yu}\ \emph {et~al.}(2012{\natexlab{a}})\citenamefont
  {Yu}, \citenamefont {Yin}, \citenamefont {Sullivan}, \citenamefont {Xia},
  \citenamefont {Huan}, \citenamefont {Paduan-Filho}, \citenamefont {{Oliveira
  Jr}}, \citenamefont {Haas}, \citenamefont {Steppke}, \citenamefont {Miclea},
  \citenamefont {Weickert}, \citenamefont {Movshovich}, \citenamefont {Mun},
  \citenamefont {Scott}, \citenamefont {Zapf},\ and\ \citenamefont
  {Roscilde}}]{YuYin_Nat_2012_DTNboseglass}%
  \BibitemOpen
  \bibfield  {author} {\bibinfo {author} {\bibfnamefont {R.}~\bibnamefont
  {Yu}}, \bibinfo {author} {\bibfnamefont {L.}~\bibnamefont {Yin}}, \bibinfo
  {author} {\bibfnamefont {N.~S.}\ \bibnamefont {Sullivan}}, \bibinfo {author}
  {\bibfnamefont {J.~S.}\ \bibnamefont {Xia}}, \bibinfo {author} {\bibfnamefont
  {C.}~\bibnamefont {Huan}}, \bibinfo {author} {\bibfnamefont {A.}~\bibnamefont
  {Paduan-Filho}}, \bibinfo {author} {\bibfnamefont {N.~F.}\ \bibnamefont
  {{Oliveira Jr}}}, \bibinfo {author} {\bibfnamefont {S.}~\bibnamefont {Haas}},
  \bibinfo {author} {\bibfnamefont {A.}~\bibnamefont {Steppke}}, \bibinfo
  {author} {\bibfnamefont {C.~F.}\ \bibnamefont {Miclea}}, \bibinfo {author}
  {\bibfnamefont {F.}~\bibnamefont {Weickert}}, \bibinfo {author}
  {\bibfnamefont {R.}~\bibnamefont {Movshovich}}, \bibinfo {author}
  {\bibfnamefont {E.-D.}\ \bibnamefont {Mun}}, \bibinfo {author} {\bibfnamefont
  {B.~L.}\ \bibnamefont {Scott}}, \bibinfo {author} {\bibfnamefont {V.~S.}\
  \bibnamefont {Zapf}}, \ and\ \bibinfo {author} {\bibfnamefont
  {T.}~\bibnamefont {Roscilde}},\ }\bibfield  {title} {\enquote {\bibinfo
  {title} {{Bose glass and Mott glass of quasiparticles in a doped quantum
  magnet}},}\ }\href {\doibase 10.1038/nature11406} {\bibfield  {journal}
  {\bibinfo  {journal} {Nature}\ }\textbf {\bibinfo {volume} {489}},\ \bibinfo
  {pages} {379} (\bibinfo {year} {2012}{\natexlab{a}})}\BibitemShut {NoStop}%
\bibitem [{\citenamefont {Zapf}\ \emph {et~al.}(2006)\citenamefont {Zapf},
  \citenamefont {Zocco}, \citenamefont {Hansen}, \citenamefont {Jaime},
  \citenamefont {Harrison}, \citenamefont {Batista}, \citenamefont
  {Kenzelmann}, \citenamefont {Niedermayer}, \citenamefont {Lacerda},\ and\
  \citenamefont {Paduan-Filho}}]{Zapf_PRL_2006_BECinDTN}%
  \BibitemOpen
  \bibfield  {author} {\bibinfo {author} {\bibfnamefont {V.~S.}\ \bibnamefont
  {Zapf}}, \bibinfo {author} {\bibfnamefont {D.}~\bibnamefont {Zocco}},
  \bibinfo {author} {\bibfnamefont {B.~R.}\ \bibnamefont {Hansen}}, \bibinfo
  {author} {\bibfnamefont {M.}~\bibnamefont {Jaime}}, \bibinfo {author}
  {\bibfnamefont {N.}~\bibnamefont {Harrison}}, \bibinfo {author}
  {\bibfnamefont {C.~D.}\ \bibnamefont {Batista}}, \bibinfo {author}
  {\bibfnamefont {M.}~\bibnamefont {Kenzelmann}}, \bibinfo {author}
  {\bibfnamefont {C.}~\bibnamefont {Niedermayer}}, \bibinfo {author}
  {\bibfnamefont {A.}~\bibnamefont {Lacerda}}, \ and\ \bibinfo {author}
  {\bibfnamefont {A.}~\bibnamefont {Paduan-Filho}},\ }\bibfield  {title}
  {\enquote {\bibinfo {title} {{Bose-Einstein Condensation of $S=1$ Nickel Spin
  Degrees of Freedom in
  ${\mathrm{NiCl}}_{2}\mathrm{\text{-}}4\mathrm{SC}({\mathrm{NH}}_{2}{)}_{2}$}},}\
  }\href {\doibase 10.1103/PhysRevLett.96.077204} {\bibfield  {journal}
  {\bibinfo  {journal} {Phys. Rev. Lett.}\ }\textbf {\bibinfo {volume} {96}},\
  \bibinfo {pages} {077204} (\bibinfo {year} {2006})}\BibitemShut {NoStop}%
\bibitem [{\citenamefont {Zvyagin}\ \emph {et~al.}(2007)\citenamefont
  {Zvyagin}, \citenamefont {Wosnitza}, \citenamefont {Batista}, \citenamefont
  {Tsukamoto}, \citenamefont {Kawashima}, \citenamefont {Krzystek},
  \citenamefont {Zapf}, \citenamefont {Jaime}, \citenamefont {Oliveira},\ and\
  \citenamefont {Paduan-Filho}}]{Zvyagin_PRL_2007_ESRinDTN}%
  \BibitemOpen
  \bibfield  {author} {\bibinfo {author} {\bibfnamefont {S.~A.}\ \bibnamefont
  {Zvyagin}}, \bibinfo {author} {\bibfnamefont {J.}~\bibnamefont {Wosnitza}},
  \bibinfo {author} {\bibfnamefont {C.~D.}\ \bibnamefont {Batista}}, \bibinfo
  {author} {\bibfnamefont {M.}~\bibnamefont {Tsukamoto}}, \bibinfo {author}
  {\bibfnamefont {N.}~\bibnamefont {Kawashima}}, \bibinfo {author}
  {\bibfnamefont {J.}~\bibnamefont {Krzystek}}, \bibinfo {author}
  {\bibfnamefont {V.~S.}\ \bibnamefont {Zapf}}, \bibinfo {author}
  {\bibfnamefont {M.}~\bibnamefont {Jaime}}, \bibinfo {author} {\bibfnamefont
  {N.~F.}\ \bibnamefont {Oliveira}}, \ and\ \bibinfo {author} {\bibfnamefont
  {A.}~\bibnamefont {Paduan-Filho}},\ }\bibfield  {title} {\enquote {\bibinfo
  {title} {{Magnetic Excitations in the Spin-1 Anisotropic Heisenberg
  Antiferromagnetic Chain System
  ${\mathrm{NiCl}}_{2}\mathrm{\text{-}}4\mathrm{SC}({\mathrm{NH}}_{2}{)}_{2}$}},}\
  }\href {\doibase 10.1103/PhysRevLett.98.047205} {\bibfield  {journal}
  {\bibinfo  {journal} {Phys. Rev. Lett.}\ }\textbf {\bibinfo {volume} {98}},\
  \bibinfo {pages} {047205} (\bibinfo {year} {2007})}\BibitemShut {NoStop}%
\bibitem [{\citenamefont {Yin}\ \emph {et~al.}(2008)\citenamefont {Yin},
  \citenamefont {Xia}, \citenamefont {Zapf}, \citenamefont {Sullivan},\ and\
  \citenamefont {Paduan-Filho}}]{YinXia_PRL_2008_DTNcritical}%
  \BibitemOpen
  \bibfield  {author} {\bibinfo {author} {\bibfnamefont {L.}~\bibnamefont
  {Yin}}, \bibinfo {author} {\bibfnamefont {J.~S.}\ \bibnamefont {Xia}},
  \bibinfo {author} {\bibfnamefont {V.~S.}\ \bibnamefont {Zapf}}, \bibinfo
  {author} {\bibfnamefont {N.~S.}\ \bibnamefont {Sullivan}}, \ and\ \bibinfo
  {author} {\bibfnamefont {A.}~\bibnamefont {Paduan-Filho}},\ }\bibfield
  {title} {\enquote {\bibinfo {title} {{Direct Measurement of the Bose-Einstein
  Condensation Universality Class in
  ${\mathrm{NiCl}}_{2}\mathrm{\text{-}}4\mathrm{SC}({\mathrm{NH}}_{2}{)}_{2}$
  at Ultralow Temperatures}},}\ }\href {\doibase
  10.1103/PhysRevLett.101.187205} {\bibfield  {journal} {\bibinfo  {journal}
  {Phys. Rev. Lett.}\ }\textbf {\bibinfo {volume} {101}},\ \bibinfo {pages}
  {187205} (\bibinfo {year} {2008})}\BibitemShut {NoStop}%
\bibitem [{\citenamefont {Blinder}\ \emph {et~al.}(2017)\citenamefont
  {Blinder}, \citenamefont {Dupont}, \citenamefont {Mukhopadhyay},
  \citenamefont {Grbi\ifmmode~\acute{c}\else \'{c}\fi{}}, \citenamefont
  {Laflorencie}, \citenamefont {Capponi}, \citenamefont {Mayaffre},
  \citenamefont {Berthier}, \citenamefont {Paduan-Filho},\ and\ \citenamefont
  {Horvati\ifmmode~\acute{c}\else \'{c}\fi{}}}]{BlinderDupont_PRB_2017_DTNNMR}%
  \BibitemOpen
  \bibfield  {author} {\bibinfo {author} {\bibfnamefont {R.}~\bibnamefont
  {Blinder}}, \bibinfo {author} {\bibfnamefont {M.}~\bibnamefont {Dupont}},
  \bibinfo {author} {\bibfnamefont {S.}~\bibnamefont {Mukhopadhyay}}, \bibinfo
  {author} {\bibfnamefont {M.~S.}\ \bibnamefont {Grbi\ifmmode~\acute{c}\else
  \'{c}\fi{}}}, \bibinfo {author} {\bibfnamefont {N.}~\bibnamefont
  {Laflorencie}}, \bibinfo {author} {\bibfnamefont {S.}~\bibnamefont
  {Capponi}}, \bibinfo {author} {\bibfnamefont {H.}~\bibnamefont {Mayaffre}},
  \bibinfo {author} {\bibfnamefont {C.}~\bibnamefont {Berthier}}, \bibinfo
  {author} {\bibfnamefont {Ar.}\ \bibnamefont {Paduan-Filho}}, \ and\ \bibinfo
  {author} {\bibfnamefont {M.}~\bibnamefont {Horvati\ifmmode~\acute{c}\else
  \'{c}\fi{}}},\ }\bibfield  {title} {\enquote {\bibinfo {title} {{Nuclear
  magnetic resonance study of the magnetic-field-induced ordered phase in the
  ${\text{NiCl}}_{2}\text{\ensuremath{-}}4\text{SC}{({\text{NH}}_{2})}_{2}$
  compound}},}\ }\href {\doibase 10.1103/PhysRevB.95.020404} {\bibfield
  {journal} {\bibinfo  {journal} {Phys. Rev. B}\ }\textbf {\bibinfo {volume}
  {95}},\ \bibinfo {pages} {020404} (\bibinfo {year} {2017})}\BibitemShut
  {NoStop}%
\bibitem [{\citenamefont {Ollivier}\ and\ \citenamefont
  {Mutka}(2011)}]{OllivierMutka_JPSJ_2011_IN5}%
  \BibitemOpen
  \bibfield  {author} {\bibinfo {author} {\bibfnamefont {J.}~\bibnamefont
  {Ollivier}}\ and\ \bibinfo {author} {\bibfnamefont {H.}~\bibnamefont
  {Mutka}},\ }\bibfield  {title} {\enquote {\bibinfo {title} {{IN5 cold neutron
  time-of-flight spectrometer, prepared to tackle single crystal
  spectroscopy}},}\ }\href {\doibase 10.1143/JPSJS.80SB.SB003} {\bibfield
  {journal} {\bibinfo  {journal} {J. Phys. Soc. Jap.}\ }\textbf {\bibinfo
  {volume} {80}},\ \bibinfo {pages} {SB003} (\bibinfo {year}
  {2011})}\BibitemShut {NoStop}%
\bibitem [{\citenamefont {Zheludev}\ and\ \citenamefont
  {Roscilde}(2013)}]{ZheludevRoscilde_CRPhysique_2013_ReviewDirtyBosons}%
  \BibitemOpen
  \bibfield  {author} {\bibinfo {author} {\bibfnamefont {A.}~\bibnamefont
  {Zheludev}}\ and\ \bibinfo {author} {\bibfnamefont {T.}~\bibnamefont
  {Roscilde}},\ }\bibfield  {title} {\enquote {\bibinfo {title} {Dirty-boson
  physics with magnetic insulators},}\ }\href {\doibase
  10.1016/j.crhy.2013.10.001} {\bibfield  {journal} {\bibinfo  {journal} {C. R.
  Physique}\ }\textbf {\bibinfo {volume} {14}},\ \bibinfo {pages} {740}
  (\bibinfo {year} {2013})}\BibitemShut {NoStop}%
\bibitem [{\citenamefont {Dupont}\ \emph {et~al.}(2017)\citenamefont {Dupont},
  \citenamefont {Capponi},\ and\ \citenamefont
  {Laflorencie}}]{DupontCapponi_PRL_2017_DTNX2BEC}%
  \BibitemOpen
  \bibfield  {author} {\bibinfo {author} {\bibfnamefont {M.}~\bibnamefont
  {Dupont}}, \bibinfo {author} {\bibfnamefont {S.}~\bibnamefont {Capponi}}, \
  and\ \bibinfo {author} {\bibfnamefont {N.}~\bibnamefont {Laflorencie}},\
  }\bibfield  {title} {\enquote {\bibinfo {title} {{Disorder-Induced Revival of
  the Bose-Einstein Condensation in
  $\mathrm{Ni}({\mathrm{Cl}}_{1\ensuremath{-}x}{\mathrm{Br}}_{x}{)}_{2}\text{\ensuremath{-}}4\mathrm{SC}({\mathrm{NH}}_{2}{)}_{2}$
  at High Magnetic Fields}},}\ }\href {\doibase 10.1103/PhysRevLett.118.067204}
  {\bibfield  {journal} {\bibinfo  {journal} {Phys. Rev. Lett.}\ }\textbf
  {\bibinfo {volume} {118}},\ \bibinfo {pages} {067204} (\bibinfo {year}
  {2017})}\BibitemShut {NoStop}%
\bibitem [{\citenamefont {Orlova}\ \emph {et~al.}(2017)\citenamefont {Orlova},
  \citenamefont {Blinder}, \citenamefont {Kermarrec}, \citenamefont {Dupont},
  \citenamefont {Laflorencie}, \citenamefont {Capponi}, \citenamefont
  {Mayaffre}, \citenamefont {Berthier}, \citenamefont {Paduan-Filho},\ and\
  \citenamefont {Horvati\'{c}}}]{OrlovaBlinder_PRL_2017_DTNXNMR}%
  \BibitemOpen
  \bibfield  {author} {\bibinfo {author} {\bibfnamefont {A.}~\bibnamefont
  {Orlova}}, \bibinfo {author} {\bibfnamefont {R.}~\bibnamefont {Blinder}},
  \bibinfo {author} {\bibfnamefont {E.}~\bibnamefont {Kermarrec}}, \bibinfo
  {author} {\bibfnamefont {M.}~\bibnamefont {Dupont}}, \bibinfo {author}
  {\bibfnamefont {N.}~\bibnamefont {Laflorencie}}, \bibinfo {author}
  {\bibfnamefont {S.}~\bibnamefont {Capponi}}, \bibinfo {author} {\bibfnamefont
  {H.}~\bibnamefont {Mayaffre}}, \bibinfo {author} {\bibfnamefont
  {C.}~\bibnamefont {Berthier}}, \bibinfo {author} {\bibfnamefont
  {A.}~\bibnamefont {Paduan-Filho}}, \ and\ \bibinfo {author} {\bibfnamefont
  {M.}~\bibnamefont {Horvati\'{c}}},\ }\bibfield  {title} {\enquote {\bibinfo
  {title} {{Nuclear Magnetic Resonance Reveals Disordered Level-Crossing
  Physics in the Bose-Glass Regime of the Br-Doped
  $\mathrm{Ni}({\mathrm{Cl}}_{1\ensuremath{-}x}{\mathrm{Br}}_{x}{)}_{2}\text{\ensuremath{-}}4\mathrm{SC}({\mathrm{NH}}_{2}{)}_{2}$
  Compound at a High Magnetic Field}},}\ }\href {\doibase
  10.1103/PhysRevLett.118.067203} {\bibfield  {journal} {\bibinfo  {journal}
  {Phys. Rev. Lett.}\ }\textbf {\bibinfo {volume} {118}},\ \bibinfo {pages}
  {067203} (\bibinfo {year} {2017})}\BibitemShut {NoStop}%
\bibitem [{\citenamefont {Wulf}(2015)}]{Wulf_2015_PhDthesis}%
  \BibitemOpen
  \bibfield  {author} {\bibinfo {author} {\bibfnamefont {E.}~\bibnamefont
  {Wulf}},\ }\href {\doibase 10.3929/ethz-a-010611031} {\emph {\bibinfo {title}
  {Experimental studies on quantum magnets in the presence of disorder}}}\
  (\bibinfo  {publisher} {PhD thesis, ETH Z\"urich},\ \bibinfo {year}
  {2015})\BibitemShut {NoStop}%
\bibitem [{Note1()}]{Note1}%
  \BibitemOpen
  \bibinfo {note} {A detailed analysis of this data will be given elsewhere
  (Mannig, Povarov \protect \emph {et al.}, in preparation).}\BibitemShut
  {Stop}%
\bibitem [{\citenamefont {Squires}(2012)}]{Squires_2012_Neutronbook}%
  \BibitemOpen
  \bibfield  {author} {\bibinfo {author} {\bibfnamefont {G.~L.}\ \bibnamefont
  {Squires}},\ }\href
  {http://www.cambridge.org/ch/academic/subjects/physics/condensed-matter-physics-nanoscience-and-mesoscopic-physics/introduction-theory-thermal-neutron-scattering-3rd-edition?format=PB}
  {\emph {\bibinfo {title} {Introduction to the Theory of Thermal Neutron
  Scattering}}}\ (\bibinfo  {publisher} {Cambridge University Press, Cambridge,
  U.K.},\ \bibinfo {year} {2012})\BibitemShut {NoStop}%
\bibitem [{\citenamefont {Tsyrulin}\ \emph {et~al.}(2013)\citenamefont
  {Tsyrulin}, \citenamefont {Batista}, \citenamefont {Zapf}, \citenamefont
  {Jaime}, \citenamefont {Hansen}, \citenamefont {Niedermayer}, \citenamefont
  {Rule}, \citenamefont {Habicht}, \citenamefont {Prokes}, \citenamefont
  {Kiefer}, \citenamefont {Ressouche}, \citenamefont {Paduan-Filho},\ and\
  \citenamefont {Kenzelmann}}]{Tsyrulin_JPCM_2013_DTNneutrons}%
  \BibitemOpen
  \bibfield  {author} {\bibinfo {author} {\bibfnamefont {N.}~\bibnamefont
  {Tsyrulin}}, \bibinfo {author} {\bibfnamefont {C.~D.}\ \bibnamefont
  {Batista}}, \bibinfo {author} {\bibfnamefont {V.~S.}\ \bibnamefont {Zapf}},
  \bibinfo {author} {\bibfnamefont {M.}~\bibnamefont {Jaime}}, \bibinfo
  {author} {\bibfnamefont {B.~R.}\ \bibnamefont {Hansen}}, \bibinfo {author}
  {\bibfnamefont {C.}~\bibnamefont {Niedermayer}}, \bibinfo {author}
  {\bibfnamefont {K.~C.}\ \bibnamefont {Rule}}, \bibinfo {author}
  {\bibfnamefont {K.}~\bibnamefont {Habicht}}, \bibinfo {author} {\bibfnamefont
  {K.}~\bibnamefont {Prokes}}, \bibinfo {author} {\bibfnamefont
  {K.}~\bibnamefont {Kiefer}}, \bibinfo {author} {\bibfnamefont
  {E.}~\bibnamefont {Ressouche}}, \bibinfo {author} {\bibfnamefont
  {A.}~\bibnamefont {Paduan-Filho}}, \ and\ \bibinfo {author} {\bibfnamefont
  {M.}~\bibnamefont {Kenzelmann}},\ }\bibfield  {title} {\enquote {\bibinfo
  {title} {{Neutron study of the magnetism in
  ${\mathrm{NiCl}}_{2}\mathrm{\text{-}}4\mathrm{SC}({\mathrm{NH}}_{2}{)}_{2}$}},}\
  }\href {\doibase 10.1088/0953-8984/25/21/216008} {\bibfield  {journal}
  {\bibinfo  {journal} {J. Phys.: Condens. Matter}\ }\textbf {\bibinfo {volume}
  {25}},\ \bibinfo {pages} {216008} (\bibinfo {year} {2013})}\BibitemShut
  {NoStop}%
\bibitem [{\citenamefont {Wulf}\ \emph {et~al.}(2013)\citenamefont {Wulf},
  \citenamefont {H\"{u}vonen}, \citenamefont {Kim}, \citenamefont
  {Paduan-Filho}, \citenamefont {Ressouche}, \citenamefont {Gvasaliya},
  \citenamefont {Zapf},\ and\ \citenamefont
  {Zheludev}}]{WulfHuvonen_PRB_2013_DTNXdiffraction}%
  \BibitemOpen
  \bibfield  {author} {\bibinfo {author} {\bibfnamefont {E.}~\bibnamefont
  {Wulf}}, \bibinfo {author} {\bibfnamefont {D.}~\bibnamefont {H\"{u}vonen}},
  \bibinfo {author} {\bibfnamefont {J.-W.}\ \bibnamefont {Kim}}, \bibinfo
  {author} {\bibfnamefont {A.}~\bibnamefont {Paduan-Filho}}, \bibinfo {author}
  {\bibfnamefont {E.}~\bibnamefont {Ressouche}}, \bibinfo {author}
  {\bibfnamefont {S.}~\bibnamefont {Gvasaliya}}, \bibinfo {author}
  {\bibfnamefont {V.}~\bibnamefont {Zapf}}, \ and\ \bibinfo {author}
  {\bibfnamefont {A.}~\bibnamefont {Zheludev}},\ }\bibfield  {title} {\enquote
  {\bibinfo {title} {Criticality in a disordered quantum antiferromagnet
  studied by neutron diffraction},}\ }\href {\doibase
  10.1103/PhysRevB.88.174418} {\bibfield  {journal} {\bibinfo  {journal} {Phys.
  Rev. B}\ }\textbf {\bibinfo {volume} {88}},\ \bibinfo {pages} {174418}
  (\bibinfo {year} {2013})}\BibitemShut {NoStop}%
\bibitem [{Note2()}]{Note2}%
  \BibitemOpen
  \bibinfo {note} {The value of $x_c\simeq 0.16$ is also consistent with
  available microscopic models for DTNX~\cite
  {DupontCapponi_PRL_2017_DTNX2BEC}: It corresponds to the percolation
  threshold of nickel sites, affected by neighboring bromine.}\BibitemShut
  {Stop}%
\bibitem [{\citenamefont {Povarov}\ \emph
  {et~al.}(2015{\natexlab{b}})\citenamefont {Povarov}, \citenamefont
  {Schmidiger}, \citenamefont {Reynolds}, \citenamefont {Bewley},\ and\
  \citenamefont {Zheludev}}]{PovarovSchmidiger_PRB_2015_DimpyScaling}%
  \BibitemOpen
  \bibfield  {author} {\bibinfo {author} {\bibfnamefont {{\relax K.
  Yu.}}~\bibnamefont {Povarov}}, \bibinfo {author} {\bibfnamefont
  {D.}~\bibnamefont {Schmidiger}}, \bibinfo {author} {\bibfnamefont
  {N.}~\bibnamefont {Reynolds}}, \bibinfo {author} {\bibfnamefont
  {R.}~\bibnamefont {Bewley}}, \ and\ \bibinfo {author} {\bibfnamefont
  {A.}~\bibnamefont {Zheludev}},\ }\bibfield  {title} {\enquote {\bibinfo
  {title} {{Scaling of temporal correlations in an attractive
  Tomonaga-Luttinger spin liquid}},}\ }\href {\doibase
  10.1103/PhysRevB.91.020406} {\bibfield  {journal} {\bibinfo  {journal} {Phys.
  Rev. B}\ }\textbf {\bibinfo {volume} {91}},\ \bibinfo {pages} {020406}
  (\bibinfo {year} {2015}{\natexlab{b}})}\BibitemShut {NoStop}%
\bibitem [{\citenamefont {H\"alg}\ \emph
  {et~al.}(2015{\natexlab{b}})\citenamefont {H\"alg}, \citenamefont
  {H\"uvonen}, \citenamefont {Butch}, \citenamefont {Demmel},\ and\
  \citenamefont {Zheludev}}]{HaelgHuvonen_PRB_2015_Dioxanescaling}%
  \BibitemOpen
  \bibfield  {author} {\bibinfo {author} {\bibfnamefont {M.}~\bibnamefont
  {H\"alg}}, \bibinfo {author} {\bibfnamefont {D.}~\bibnamefont {H\"uvonen}},
  \bibinfo {author} {\bibfnamefont {N.~P.}\ \bibnamefont {Butch}}, \bibinfo
  {author} {\bibfnamefont {F.}~\bibnamefont {Demmel}}, \ and\ \bibinfo {author}
  {\bibfnamefont {A.}~\bibnamefont {Zheludev}},\ }\bibfield  {title} {\enquote
  {\bibinfo {title} {{Finite-temperature scaling of spin correlations in a
  partially magnetized Heisenberg $S=\frac{1}{2}$ chain}},}\ }\href {\doibase
  10.1103/PhysRevB.92.104416} {\bibfield  {journal} {\bibinfo  {journal} {Phys.
  Rev. B}\ }\textbf {\bibinfo {volume} {92}},\ \bibinfo {pages} {104416}
  (\bibinfo {year} {2015}{\natexlab{b}})}\BibitemShut {NoStop}%
\bibitem [{\citenamefont {Zhang}\ \emph {et~al.}(2013)\citenamefont {Zhang},
  \citenamefont {Wierschem}, \citenamefont {Yap}, \citenamefont {Kato},
  \citenamefont {Batista},\ and\ \citenamefont
  {Sengupta}}]{ZhangWierschem_PRB_2013_DTNdispersion}%
  \BibitemOpen
  \bibfield  {author} {\bibinfo {author} {\bibfnamefont {Z.}~\bibnamefont
  {Zhang}}, \bibinfo {author} {\bibfnamefont {K.}~\bibnamefont {Wierschem}},
  \bibinfo {author} {\bibfnamefont {I.}~\bibnamefont {Yap}}, \bibinfo {author}
  {\bibfnamefont {Y.}~\bibnamefont {Kato}}, \bibinfo {author} {\bibfnamefont
  {C.~D.}\ \bibnamefont {Batista}}, \ and\ \bibinfo {author} {\bibfnamefont
  {P.}~\bibnamefont {Sengupta}},\ }\bibfield  {title} {\enquote {\bibinfo
  {title} {Phase diagram and magnetic excitations of anisotropic spin-one
  magnets},}\ }\href {\doibase 10.1103/PhysRevB.87.174405} {\bibfield
  {journal} {\bibinfo  {journal} {Phys. Rev. B}\ }\textbf {\bibinfo {volume}
  {87}},\ \bibinfo {pages} {174405} (\bibinfo {year} {2013})}\BibitemShut
  {NoStop}%
\bibitem [{\citenamefont {Wulf}\ \emph {et~al.}(2015)\citenamefont {Wulf},
  \citenamefont {H\"{u}vonen}, \citenamefont {Sch\"{o}nemann}, \citenamefont
  {K\"{u}hne}, \citenamefont {Herrmannsd\"{o}rfer}, \citenamefont {Glavatskyy},
  \citenamefont {Gerischer}, \citenamefont {Kiefer}, \citenamefont
  {Gvasaliya},\ and\ \citenamefont
  {Zheludev}}]{WulfHuvonen_PRB_2015_DTNIntrinsicBroadening}%
  \BibitemOpen
  \bibfield  {author} {\bibinfo {author} {\bibfnamefont {E.}~\bibnamefont
  {Wulf}}, \bibinfo {author} {\bibfnamefont {D.}~\bibnamefont {H\"{u}vonen}},
  \bibinfo {author} {\bibfnamefont {R.}~\bibnamefont {Sch\"{o}nemann}},
  \bibinfo {author} {\bibfnamefont {H.}~\bibnamefont {K\"{u}hne}}, \bibinfo
  {author} {\bibfnamefont {T.}~\bibnamefont {Herrmannsd\"{o}rfer}}, \bibinfo
  {author} {\bibfnamefont {I.}~\bibnamefont {Glavatskyy}}, \bibinfo {author}
  {\bibfnamefont {S.}~\bibnamefont {Gerischer}}, \bibinfo {author}
  {\bibfnamefont {K.}~\bibnamefont {Kiefer}}, \bibinfo {author} {\bibfnamefont
  {S.}~\bibnamefont {Gvasaliya}}, \ and\ \bibinfo {author} {\bibfnamefont
  {A.}~\bibnamefont {Zheludev}},\ }\bibfield  {title} {\enquote {\bibinfo
  {title} {{Critical exponents and intrinsic broadening of the field-induced
  transition in
  ${\mathrm{NiCl}}_{2}\mathrm{\text{-}}4\mathrm{SC}({\mathrm{NH}}_{2}{)}_{2}$}},}\
  }\href {\doibase 10.1103/PhysRevB.91.014406} {\bibfield  {journal} {\bibinfo
  {journal} {Phys. Rev. B}\ }\textbf {\bibinfo {volume} {91}},\ \bibinfo
  {pages} {014406} (\bibinfo {year} {2015})}\BibitemShut {NoStop}%
\bibitem [{\citenamefont {Fisher}\ \emph {et~al.}(1989)\citenamefont {Fisher},
  \citenamefont {Weichman}, \citenamefont {Grinstein},\ and\ \citenamefont
  {Fisher}}]{Fisher_PRB_1989_Boseglass}%
  \BibitemOpen
  \bibfield  {author} {\bibinfo {author} {\bibfnamefont {M.~P.~A.}\
  \bibnamefont {Fisher}}, \bibinfo {author} {\bibfnamefont {P.~B.}\
  \bibnamefont {Weichman}}, \bibinfo {author} {\bibfnamefont {G.}~\bibnamefont
  {Grinstein}}, \ and\ \bibinfo {author} {\bibfnamefont {D.~S.}\ \bibnamefont
  {Fisher}},\ }\bibfield  {title} {\enquote {\bibinfo {title} {Boson
  localization and the superfluid-insulator transition},}\ }\href {\doibase
  10.1103/PhysRevB.40.546} {\bibfield  {journal} {\bibinfo  {journal} {Phys.
  Rev. B}\ }\textbf {\bibinfo {volume} {40}},\ \bibinfo {pages} {546--570}
  (\bibinfo {year} {1989})}\BibitemShut {NoStop}%
\bibitem [{\citenamefont {Yu}\ \emph {et~al.}(2012{\natexlab{b}})\citenamefont
  {Yu}, \citenamefont {Miclea}, \citenamefont {Weickert}, \citenamefont
  {Movshovich}, \citenamefont {Paduan-Filho}, \citenamefont {Zapf},\ and\
  \citenamefont {Roscilde}}]{YuMiclea_PRB_2012_DTNXexponents}%
  \BibitemOpen
  \bibfield  {author} {\bibinfo {author} {\bibfnamefont {R.}~\bibnamefont
  {Yu}}, \bibinfo {author} {\bibfnamefont {C.~F.}\ \bibnamefont {Miclea}},
  \bibinfo {author} {\bibfnamefont {F.}~\bibnamefont {Weickert}}, \bibinfo
  {author} {\bibfnamefont {R.}~\bibnamefont {Movshovich}}, \bibinfo {author}
  {\bibfnamefont {A.}~\bibnamefont {Paduan-Filho}}, \bibinfo {author}
  {\bibfnamefont {V.~S.}\ \bibnamefont {Zapf}}, \ and\ \bibinfo {author}
  {\bibfnamefont {T.}~\bibnamefont {Roscilde}},\ }\bibfield  {title} {\enquote
  {\bibinfo {title} {{Quantum critical scaling at a Bose-glass/superfluid
  transition: Theory and experiment for a model quantum magnet}},}\ }\href
  {\doibase 10.1103/PhysRevB.86.134421} {\bibfield  {journal} {\bibinfo
  {journal} {Phys. Rev. B}\ }\textbf {\bibinfo {volume} {86}},\ \bibinfo
  {pages} {134421} (\bibinfo {year} {2012}{\natexlab{b}})}\BibitemShut
  {NoStop}%
\bibitem [{\citenamefont {Harris}(1974)}]{Harris_JPhysC_1974_HarrisCriter}%
  \BibitemOpen
  \bibfield  {author} {\bibinfo {author} {\bibfnamefont {A.~B.}\ \bibnamefont
  {Harris}},\ }\bibfield  {title} {\enquote {\bibinfo {title} {{Effect of
  random defects on the critical behavior of Ising models}},}\ }\href {\doibase
  10.1088/0022-3719/7/9/009} {\bibfield  {journal} {\bibinfo  {journal} {J.
  Phys. C: Solid State Phys.}\ }\textbf {\bibinfo {volume} {7}},\ \bibinfo
  {pages} {1671} (\bibinfo {year} {1974})}\BibitemShut {NoStop}%
\bibitem [{\citenamefont {Yao}\ \emph {et~al.}(2014)\citenamefont {Yao},
  \citenamefont {{da Costa}}, \citenamefont {Kiselev},\ and\ \citenamefont
  {Prokof'ev}}]{ZhiyuanProkofev_PRL_2014_BGexponent}%
  \BibitemOpen
  \bibfield  {author} {\bibinfo {author} {\bibfnamefont {Z.}~\bibnamefont
  {Yao}}, \bibinfo {author} {\bibfnamefont {K.~P.~C.}\ \bibnamefont {{da
  Costa}}}, \bibinfo {author} {\bibfnamefont {M.}~\bibnamefont {Kiselev}}, \
  and\ \bibinfo {author} {\bibfnamefont {N.}~\bibnamefont {Prokof'ev}},\
  }\bibfield  {title} {\enquote {\bibinfo {title} {{Critical Exponents of the
  Superfluid--Bose-Glass Transition in Three Dimensions}},}\ }\href {\doibase
  10.1103/PhysRevLett.112.225301} {\bibfield  {journal} {\bibinfo  {journal}
  {Phys. Rev. Lett.}\ }\textbf {\bibinfo {volume} {112}},\ \bibinfo {pages}
  {225301} (\bibinfo {year} {2014})}\BibitemShut {NoStop}%
\bibitem [{\citenamefont {Vojta}(2013)}]{Vojta_PRL_2013_InGap}%
  \BibitemOpen
  \bibfield  {author} {\bibinfo {author} {\bibfnamefont {M.}~\bibnamefont
  {Vojta}},\ }\bibfield  {title} {\enquote {\bibinfo {title} {Excitation
  spectra of disordered dimer magnets near quantum criticality},}\ }\href
  {\doibase 10.1103/PhysRevLett.111.097202} {\bibfield  {journal} {\bibinfo
  {journal} {Phys. Rev. Lett.}\ }\textbf {\bibinfo {volume} {111}},\ \bibinfo
  {pages} {097202} (\bibinfo {year} {2013})}\BibitemShut {NoStop}%
\end{thebibliography}%

\section{Supplemental material}

\subsection{Ordered moment estimate}

\section{Ordered moment estimate}

\begin{figure}
  \includegraphics[width=0.5\textwidth]{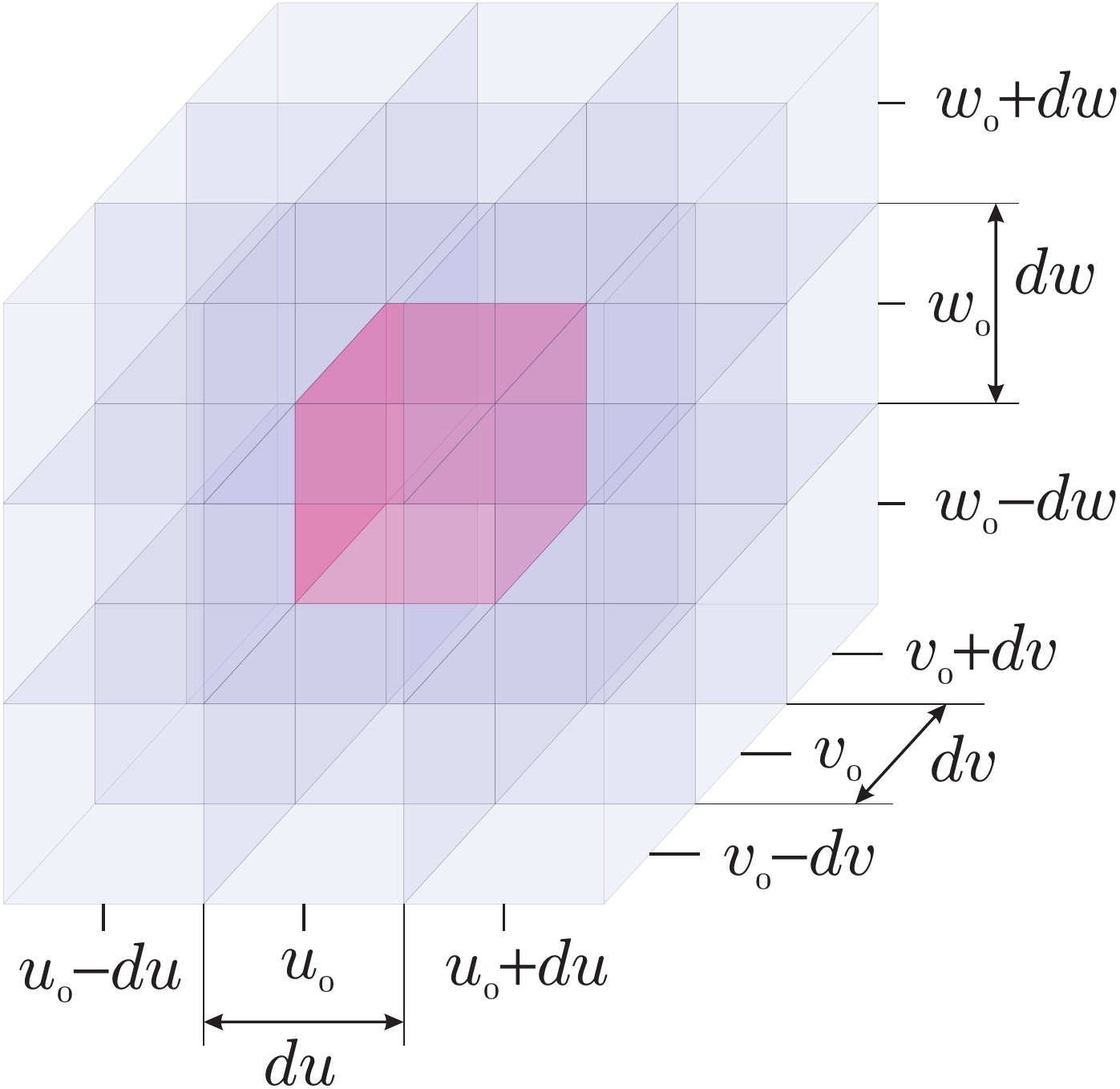}\\
  \caption{
  The Bragg peak intensity  determination in the time-of-flight data. The coordinate system
  of three orthogonal vectors is defined as $\mathbf{u}=(h,~h,~0)$,
  $\mathbf{v}=(0,~0,~l)$ and $\mathbf{w}=(-h,~h,~0)$. We determine the
  total
  intensity in the central ``red'' block (voxel) of the reciprocal space
  containing the Bragg peak at $(u_{0},~v_{0},~w_{0})$ [corresponding to $(h_{0},~h_{0},~l_{0})$
  in the standard r.l.u. notation] and the average of the intensities in
  the neighboring blocks is taken as the background.
  }
  \label{FIG:hypercube}
\end{figure}

\begin{figure}
    \includegraphics[width=0.5\textwidth]{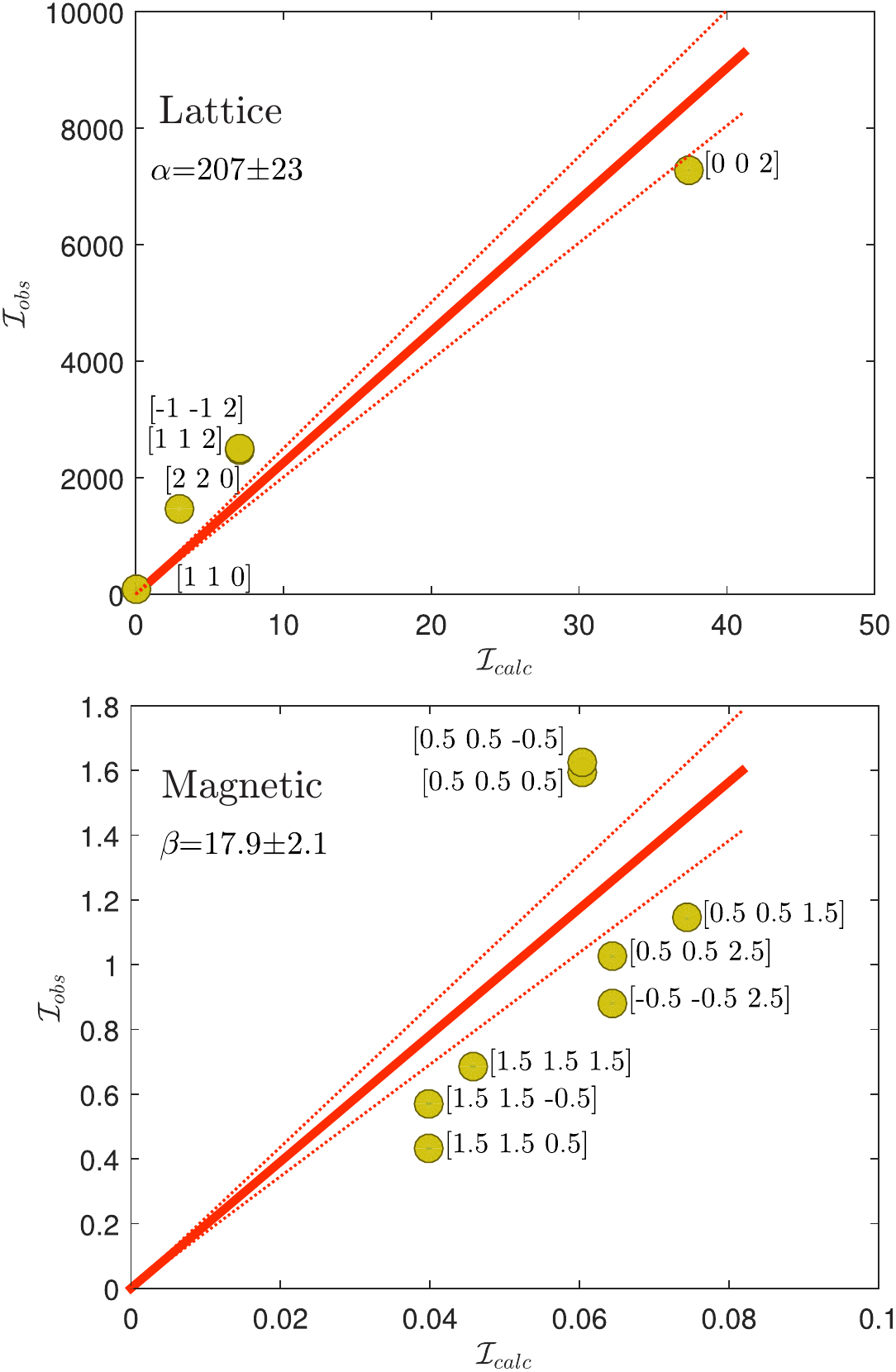}\\
    \caption{Observed versus calculated intensities for the lattice (top) and magnetic (bottom) Bragg peaks. Lines show the best linear fit with
    the tolerance intervals.}
    \label{FIG:peaks_ii}
\end{figure}

Estimate of the ordered moment was done according to the elastic
scattering theory, as given in the textbook by
Squires~\cite{Squires_2012_Neutronbook}. This analysis is based on
the fact that the intensities of lattice Bragg peaks from a given
sample on a given instrument can be used to calibrate out the
unknown prefactor, relating the measured ``arbitrary units'' to the
absolute units of scattering crossection. Then, in turn, with this
calibration the ordered magnetic moment can be extracted from the
magnetic Bragg peak intensities.

As we deal with a time-of-flight dataset, we have to extract the
peak intensities in an unconventional manner. The spectrometer has a
discrete detector bank, and the sample rotation angles in the
experiment are rather discrete too (in comparison to a dedicated
diffraction experiment). The discrete structure of the
time-of-flight data lacking momentum resolution does not allow one
to meaningfully plot the peak in a conventional ``rocking curve''
manner~\cite{Squires_2012_Neutronbook}. Instead one has to work with
the voxels of reciprocal space. The intensity within a given voxel
is the result of statistical treatment of many events on many
detectors, and there is no guarantee that the Bragg condition was
precisely matched for a detector angle and sample rotation angle. As
the result of the discreteness, the intensities in the Bragg
scattering related voxels may experience some random modulation in
the time-of-flight dataset. This makes the analysis below just a
\emph{crude estimate} of the ordered magnetic moment.

For the description of the scattering data it is convenient to use
the scattering axes basis: $\mathbf{u}=(h,~h,~0)$,
  $\mathbf{v}=(0,~0,~l)$ and $\mathbf{w}=(-h,~h,~0)$. As the first step we have found the
intensities of the peaks. For this we took the following approach
(shown in Fig.~\ref{FIG:hypercube}): for each peak we have
identified the rectangular block (voxel) of reciprocal space, fully
enclosing it. The coordinates of the block center are
$(u_{0},v_{0},w_{0})$ and its dimensions are $(du,dv,dw)$ in
$\mathbf{u}$, $\mathbf{v}$ and $\mathbf{w}$ coordinate system. The
total intensity (integrated in $\hbar\omega$ from $-0.1$ to
$0.1$~meV) in the so defined block is the sum of the peak total
intensity and the background. For the background estimate we took
the average of intensity in 26 neighboring blocks of the same size,
located at $(u_{0}\pm du,v_{0}\pm dv,w_{0}\pm dw)$. After the
intensities of the Bragg peaks are obtained, they can be compared to
the theory predictions. For the lattice Bragg peaks the intensity
is:

\begin{equation}\label{EQ:I_latt}
    \mathcal{I}_{L}(\mathbf{Q})=I_{0}\frac{(2\pi)^{3}}{V_{0}}N\left|\mathcal{F}_L(\mathbf{Q})\right|^{2},
\end{equation}

where $V_{0}$ is the unit cell volume, $N$ is the number of unit
cells in the sample, and $I_{0}$ is the unknown instrumental
coefficient (as no absolute calibration for the scattering intensity
was performed). The last term is the lattice cell structure factor:

\begin{equation}\label{EQ:I_lattFF}
    \mathcal{F}_L(\mathbf{Q})=\sum\limits^{\text{all c.c.}}_{\mathbf{r}_{j}}b_{j}e^{i(\mathbf{Q}\cdot
    \mathbf{r}_{j})}.
\end{equation}

 The vectors $\mathbf{r}_{j}$ are the positions of the atoms within the unit
 cell and $b_{j}$ are the corresponding scattering length
 parameters. The summation goes through all the atoms within the unit
 cell. The equations~(\ref{EQ:I_latt},\ref{EQ:I_lattFF}) may be reduced to
 a simpler form of relation between the observed and calculated
 intensities $ \mathcal{I}_{L}(\mathbf{Q})=\alpha
\mathcal{I}_{L}^{calc}(\mathbf{Q})$, where $\alpha=I_{0}(2\pi)^{3}N$
is the parameter of interest, which needs to be ``calibrated''.

For the magnetic elastic scattering the intensity is:

\begin{align}\label{EQ:I_mag}
    \mathcal{I}_{M}(\mathbf{Q})=I_{0}\frac{(2\pi)^{3}}{8V_{0}}\frac{N}{8}(\gamma r_{0})^{2}\left|\mathcal{F}_M(\mathbf{Q})\right|^{2}\times \\
    (1-(\mathbf{\widehat{q}\cdot
    \widehat{s}})^{2})\left(\frac{g}{2}\right)^{2}\aver{S}^{2}.
\end{align}

Here $I_{0}$, $N$, and $V_{0}$ are the same as in
Eq.~(\ref{EQ:I_latt}). The magnetic unit cell is $2^3$ times bigger
than the crystal one, and hence there is a prefactor of 8 appearing
twice --- for the cell volume and for the number of cells. The
prefactor $(\gamma r_{0})^{2}=29.06$~fm$^2$ is the universal
constant. There is also the magnetic $g$ factor of the Ni$^{2+}$ ion
along the ordered moment direction $\mathbf{\widehat{s}}$. We assume
this direction to be $\mathbf{\widehat{s}}=(1,0,0)$, and hence
$g\simeq2.3$. The last few terms are the polarization factor
(dependent on the angle between $\mathbf{\widehat{s}}$ and
scattering momentum direction
$\mathbf{\widehat{q}}=\mathbf{Q}/|Q|$), product of magnetic
form-factor and magnetic cell structure factor, and the quantity of
our main interest
--- the ordered moment squared $\aver{S}^{2}$. The product of magnetic form-factor and
magnetic cell structure factor, in turn, is expressed as

\begin{equation}\label{EQ:I_magFF}
    \mathcal{F}_M(\mathbf{Q})=F(Q)\sum\limits^{\text{Ni m.c.}}_{\mathbf{r}_{j}}\sigma_{j}e^{i(\mathbf{Q}\cdot
    \mathbf{r}_{j})}.
\end{equation}

Like in Eq.~(\ref{EQ:I_lattFF}) there is a summation over the atoms
in the unit cell. However, the difference is that now only the
magnetic ions are considered and the summation goes through the
magnetic unit cell, which is eight times bigger. There is a factor
$\sigma_{j}=\pm1$ accounting for the staggered magnetic moment, and
$F(Q)$ is the magnetic form factor of the Ni$^{2+}$ ion.

Note, that in Eqs.~(\ref{EQ:I_lattFF},\ref{EQ:I_magFF}) we have
neglected the Debye--Waller factors, related to the atomic
displacements form the equilibrium positions. At very low
temperatures these displacements are small and can be disregarded,
as we work with minor momentum transfers.

Again, we can express the relation between the observed and expected
intensities as $ \mathcal{I}_{M}(\mathbf{Q})=\beta
\mathcal{I}_{M}^{calc}(\mathbf{Q})$, with the prefactor
$\beta=I_{0}(2\pi)^{3}N\aver{S}^{2}$. Then our ordered moment value
is expressed as $\aver{S}=\sqrt{\beta/\alpha}$ in units of
$\mu_{B}$.

Comparing the observed and calculated intensities (see
Fig.~\ref{FIG:peaks_ii}) we yield $\alpha=207\pm23$ and
$\beta=18\pm2$. Hence, $\aver{S}=0.29\pm0.03~\mu_{B}$, which is
$15\pm2$\% of the full nickel ion magnetic moment. Again, we would
like to reiterate that due to the coarse nature of the dataset the
analysis above should be seen only as a crude estimate of $\langle S
\rangle$. It also is based on the (very reasonable) assumption that
the structure of the ordered phase is identical to the field-induced
case analyzed in detail by Tsyrulin~\emph{et
al.}~\cite{Tsyrulin_JPCM_2013_DTNneutrons}, and in fact does not
contain an independent attempt to solve the magnetic structure.
Finally, the given error bars are of purely statistical nature and
may not reflect a possible systematic error present due to a coarse
dataset.
\end{document}